\documentclass[twocolumn,english,pra,aps,superscriptaddress]{revtex4-1}
\usepackage[T1]{fontenc}
\usepackage{color}
\usepackage{babel}
\usepackage{mathtools}
\usepackage{bm}
\usepackage{graphicx}
\usepackage{esint}
\usepackage{amsfonts}
\usepackage{graphicx}
\usepackage{float}
\usepackage{subfigure}
\usepackage{amsthm}

\usepackage[usenames,dvipsnames,svgnames,table]{xcolor}
\usepackage[unicode=true,pdfusetitle,
bookmarks=true,bookmarksnumbered=false,bookmarksopen=false,
breaklinks=true,pdfborder={0 0 0},backref=false,colorlinks=true]{hyperref}
\hypersetup{linkcolor=NavyBlue,urlcolor=NavyBlue,citecolor=NavyBlue}
\usepackage{breakurl}

\newtheorem{prop}{Proposition}
\newtheorem{theorem}{Theorem}
\newtheorem{corollary}{Corollary}

\newtheorem{definition}{Definition}

\begin{document}

\title{Operational definition of a quantum speed limit}

\author{Yanyan Shao}
\affiliation{MOE Key Laboratory of Fundamental Physical Quantities
Measurement \& Hubei Key Laboratory of Gravitation and Quantum Physics,
PGMF and School of Physics, Huazhong University of Science and Technology,
Wuhan 430074, P. R. China}

\author{Bo Liu}
\affiliation{Beijing Computational Science Research Center, Beijing 100193, China}

\author{Mao Zhang}
\affiliation{MOE Key Laboratory of Fundamental Physical Quantities
Measurement \& Hubei Key Laboratory of Gravitation and Quantum Physics,
PGMF and School of Physics, Huazhong University of Science and Technology,
Wuhan 430074, P. R. China}

\author{Haidong Yuan}
\affiliation{Department of Mechanical and Automation Engineering, The Chinese
University of Hong Kong, Shatin, Hong Kong}

\author{Jing Liu}
\email{liujingphys@hust.edu.cn}
\affiliation{MOE Key Laboratory of Fundamental Physical Quantities
Measurement \& Hubei Key Laboratory of Gravitation and Quantum Physics,
PGMF and School of Physics, Huazhong University of Science and Technology,
Wuhan 430074, P. R. China}

\begin{abstract}
The quantum speed limit is a fundamental concept in quantum mechanics, which aims at
finding the minimum time scale or the maximum dynamical speed for some fixed
targets. In a large number of studies in this field, the construction of valid
bounds for the evolution time is always the core mission, yet the physics behind
it and some fundamental questions like which states can really fulfill the target,
are ignored. Understanding the physics behind the bounds is at least as
important as constructing attainable bounds. Here we provide an operational
approach for the definition of quantum speed limit, which utilizes the set
of states that can fulfill the target to define the speed limit. Its
performances in various scenarios have been investigated. For time-independent
Hamiltonians, it is inverse-proportional to the difference between the highest
and lowest energies. The fact that its attainability does not require a zero
ground-state energy suggests it can be used as an indicator of quantum phase
transitions. For time-dependent Hamiltonians, it is shown that contrary
to the results given by existing bounds, the true speed limit should be
independent of the time. Moreover, in the case of spontaneous emission,
we find a counterintuitive phenomenon that a lousy purity can benefit
the reduction of the quantum speed limit.
\end{abstract}

\maketitle

\section{Introduction}

Coherence and entanglement are important resources in quantum technology,
especially in quantum information processing, quantum computation~\cite{Chitambar2019}
and quantum metrology~\cite{Giovannetti2011,Toth2014}. However, the existence of
decoherence limits the lifetime of these quantum resources, and now is a major
obstacle for the development of quantum computers. Extending the coherent time and
reducing the operation time with bounded energies are two common
methods in general for this problem. To reduce the time for performing a quantum
gate, the system needs to evolve as fast as possible, and the shortest time for
performing a quantum operation or evolving a state to a target state is now
referred to as the quantum speed limit (QSL).

The QSL has now been broadly used to characterize quantum dynamics~\cite{Mandelstam1945,Margolus1998,
Giovannetti2004,Levitin2009,Zhang2014,Pires2016,Marvian2016,Deffner2017,Epstein2017,
Campaioli2018,Bukov2019,Wu2018,Liu2015,Ashhab2012,Zhang2015}. Specifically,
they have found applications in open quantum systems~\cite{Taddei2013,Campo2013,
Deffner2013,Sun2015,Marvian2015,Mirkin2016,Campo2019,Campaioli2019}, e.g.,
in the identification of decoherence times~\cite{Chenu17,Beau17b}, as well
as in quantum metrology~\cite{Giovannetti2003,Giovannetti2006,Beau17a}, quantum
control~\cite{Caneva2009,Hegerfeldt2013,Funo17,Campbell2017,Poggi2019}, and quantum
information processings like the preparation of quantum states~\cite{Girolami2019}.
They have also been studied in nonequilibrium dynamics~\cite{Cai2017},
relativistic dynamics~\cite{Villamizar2015}, and non-Hermitian systems~\cite{Sun2019}.
The recent introduction of speed limits in classical systems~\cite{Margolus11,Shanahan2018,Okuyama2018}
suggests a unifying framework of both quantum and classical bounds using information
geometry~\cite{Amari16}. Novel numerical methods like machine learning~\cite{Yung2018}
have also been applied in the study of the QSL. A thorough review on the recent
development of the QSL can be found in Ref.~\cite{Deffner2017}.

For a pure state under unitary evolution, the evolved state
$|\psi(t)\rangle=\exp(-iHt)|\psi(0)\rangle$, where $H$ is a time-independent
Hamiltonian of the system, $|\psi(0)\rangle$ is the initial state and $t$ is the
evolved time. Here and in the following, $\hbar$ is set to be 1. The most well-known
scenario for the QSL is to evolve a pure state to its orthogonal state. In this case,
the first bound for evolution time is $\tau_{\mathrm{MT}}=\pi/(2\Delta H)$,
where $\Delta H:=\sqrt{\langle H^{2}\rangle-\langle H\rangle^{2}}$ is the standard
deviation of the Hamiltonian with $\langle\cdot\rangle$ the expected value. This
bound was given by Mandelstam and Tamm in 1945~\cite{Mandelstam1945}, known as the
MT bound today. Latter in 1998, Margolus and Levitin~\cite{Margolus1998}
provided another bound for this scenario $\tau_{\mathrm{ML}}=\pi/(2 \langle H\rangle)$,
which is known as the ML bound now. In 2009, Levitin and Toffoli~\cite{Levitin2009}
proved that the combined bound of $\tau_{\mathrm{MT}}$ and $\tau_{\mathrm{ML}}$
is tight by assuming the ground energy is zero. However, this bound can only be
attained by two-level systems with the specific states $\frac{1}{\sqrt{2}}(|E_{0}\rangle
+e^{i\phi}|E_{1}\rangle)$ ($|E_{0}\rangle$, $|E_{1}\rangle$ are
the energy eigenstates and $\phi\in[0,2\pi]$ is a relative phase)~\cite{Deffner2017}.
For a more general target, this bound was numerically extended to
$\tau_{\mathrm{C}}=\max\left\{\frac{\mathcal{A}}{\Delta H},\frac{2\mathcal{A}^{2}}
{\pi\langle H\rangle}\right\}$ by Giovannetti, Lloyd and Maccone~\cite{Giovannetti2003,Giovannetti2004},
with $\mathcal{A}=\arccos f$ the Bures angle, as well as the target angle, in
this equation. $f=\mathrm{Tr}\sqrt{\sqrt{\rho_{0}}\rho_{1}\sqrt{\rho_{0}}}$ is
the fidelity between two quantum states $\rho_{0}$ and $\rho_{1}$.

Another well-used method for the construction of the QSL is the geometric
approach, which utilizes the metrics and geodesic lines in some differential
manifolds. One such example is the quantum Fisher information based on the
symmetric logarithmic derivative, which is
proportional to the Fubini-Study and Bures metrics for pure and mixed
states~\cite{Liu2020}. In 2013, Taddei \emph{et al.}~\cite{Taddei2013} used it
to construct an inequality for the QSL
$\mathcal{A}\leq \int^{t}_{0} \frac{1}{2}\sqrt{F(t^{\prime})}\mathrm{d}t^{\prime}$,
where $F(t)$ is the quantum Fisher information for the time $t$. The squared
infinitesimal distance then reads $\mathrm{d} s^2=\sum_{\mu\nu}g_{\mu\nu}\mathrm{d}
\lambda_{\mu}\mathrm{d}\lambda_{\nu}$. In the case that $F$ is independent of time,
an explicit expression of QSL can be obtained as
$\tau_{\mathrm{F}}=2\mathcal{A}/\sqrt{F}$. Similarly to the previous mentioned
tools, $\tau_{\mathrm{F}}$ are not attainable for mixed states and high-level
systems. In around 2016, Mondal \emph{et al.}~\cite{Mondal2016} extended the
result to the Wigner–Yanase skew information and connected the QSL with the
quantum coherence, and in the mean time, Pires \emph{et al.}~\cite{Pires2016}
extended this result to a family of contractive Riemannian metrics~ (also known
as a family of quantum Fisher information in some literatures)~\cite{Petz1996}.
For a density matrix $\rho$ which is a function of a set of parameters
$\{\lambda_{\mu}\}$, this family of metrics is of the form $g_{\mu\nu}=\frac{1}{4}\mathrm{Tr}
[\partial_{\lambda_{\mu}}\rho\mathbf{K}^{-1} (\partial_{\lambda_{\nu}}\rho)]$,
where $\mathbf{K}(\cdot)$ is a superoperator defined by
$\mathbf{K}(\cdot)=h(\mathbf{L}\mathbf{R}^{-1})\mathbf{R}(\cdot)$
with $\mathbf{L}$ ($\mathbf{R}$) also a superoperator defined by $\mathbf{L}(A)=A\rho$
($\mathbf{R}(A)=\rho A$). $h(\cdot)$ here is called the Morozova-\v{C}encov function,
which satisfies operator monotone ($h(A)\geq h(B)$ for $A\geq B$), self-inverse
($h(x)=xh(1/x)$) and normalization ($h(1)=1$). Assuming all the parameters in
$\{\lambda_{\mu}\}$ are dependent on time, the geodesic line $\mathcal{L}$ between
the initial and evolved states satisfies
\begin{equation}
\mathcal{L} \leq \int^{t}_0\frac{\mathrm{d} s}{\mathrm{d}t'}\mathrm{d}t'
=\int^{t}_0 \sqrt{\sum_{\mu\nu}g_{\mu\nu} \frac{\mathrm{d}\lambda_{\mu}}{\mathrm{d}t'}
\frac{\mathrm{d}\lambda_{\nu}}{\mathrm{d}t'} } \mathrm{d} t'.
\end{equation}

%========================Figure==========================
\begin{figure}[tp]
\centering
\includegraphics[width=8cm]{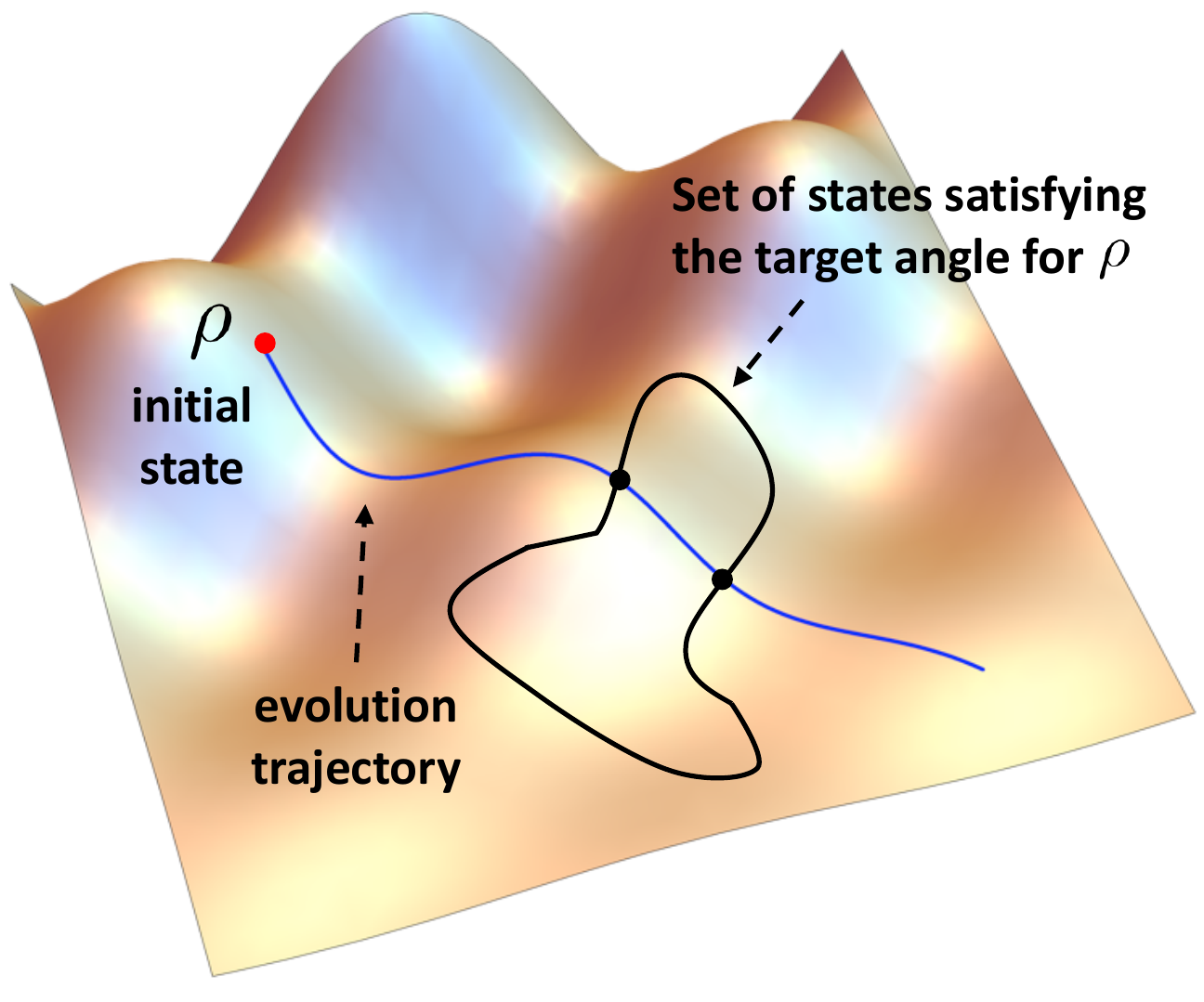}
\caption{(Color online) Dynamical trajectory of a quantum state $\rho$.
Only one trajectory (blue line) exists for a non-cotrolled fixed Hamiltonian
with a fixed decoherence. The states satisfying the target angle are hence
the intersections between the trajectory and the set of states satisfying the
target angle (black line), which is determined by the trajectory itself.
Therefore, the QSL should not be a function of time in these cases.}
\label{fig:trajectory}
\end{figure}
%========================================================

Bures angle is not the only tool to define the target angle in the studies of QSL.
For example, in 2013 del Campo \emph{et al.}~\cite{Campo2013} used the relative
purity and Campaioli \emph{et al.}~\cite{Campaioli2018} further used its angle to
define the target angle. Other types of fidelity are also considered~\cite{Sun2015,Zhang2018}.
Bloch vector is another well-used geometric representation of quantum states in
quantum mechanics, and the angle between the Bloch vectors provides another tool
to define the target angle~\cite{Campaioli2018,Campaioli2019}. Considering the
unitary evolution, Campaioli \emph{et al.}~\cite{Campaioli2018} provided an
alternative inequality for the QSL as
\begin{equation}
t\geq \tau_{\mathrm{B}} = \frac{\Theta}{Q},
\end{equation}
where $\Theta$ is the target angle defined via the Bloch vectors and
\begin{equation}
Q=\frac{1}{t}\int^{t}_{0}\sqrt{\frac{2\mathrm{Tr}(\rho^2 H^2-\rho H\rho H)}
{\mathrm{Tr}(\rho^2)-1/N}}\mathrm{d}t',
\label{eq:tauB_Q}
\end{equation}
with $N$ the dimension of $\rho$.

In most theories in regard to the QSL, an explicit inequality with respect to the time
is hard to obtain since it usually involves an integral which cannot be solved
analytically, especially in the case of time-dependent Hamiltonians. The common
method to deal with it is to formally add $t$ and $1/t$ in front of the integral
simultaneously and treat $1/t$ and the integral together as an expected value of
some quantity with respect to $t$. For example, in the inequality $\mathcal{L}\leq
\int^{t}_0 X(t')\mathrm{d}t'$, one can obtain a formal inequality on $t$
as $t\geq \mathcal{L}/\bar{X}(t)$ with $\bar{X}(t)=\frac{1}{t}\int^{t}_0 X(t')\mathrm{d}t'$
the average value with respect to time. The major problem of this formal solution
is that $\bar{X}(t)$ is a function of time in most cases, indicating the obtained
bound will change for different choice of time. However, this result does not
reflect the physics correctly. In the case of a non-controlled fixed Hamiltonian,
the trajectory of evolution in state space is fixed for a fixed decoherence mode
and strength, whether the Hamiltonian is time-dependent or not. This is due
to the fact that the solutions of states in a fixed differential equation is unique.

This can also be understood from the perspective of physics. Consider the unitary
evolution for a specific initial state $\rho$ with a time-dependent Hamiltonian.
The dynamical operator is $U(t_1)=\exp(-i\mathcal{T}\int^{t_1}_{0}H(t)\mathrm{d}t)$
with $\mathcal{T}$ the time-ordering operator. For a non-controlled Hamiltonian,
$U(t_1)$ relies on $t_1$, not $t$, indicating that $U$ is fixed for a fixed $t_1$.
This fact means the dynamical trajectory (blue line in Fig.~\ref{fig:trajectory})
in the state space for $\rho$ is fixed. In the meantime, the set of states
satisfying the target angle for $\rho$ is also fixed (black line in Fig.~\ref{fig:trajectory}).
Therefore, the states that can reach the target angle on the trajectory are the
intersections between the blue and black lines, which is actually determined by
the trajectory itself. Then the evolution time to reach the target angle for
$\rho$ is fixed in this case due to the fact that the trajectory is fixed.
In a word, this evolution time is determined by the other parameters (apart from
the evolution time) in the Hamiltonian and dissipative modes in the case of open
systems, rather than the time $t$. Hence, the QSL should not be dependent on the
time either. Most of the current theoretical tools cannot reveal this fact,
especially for the time-dependent Hamiltonians. New approaches are still in
need in this field to reveal the true physics behind the QSL. This is a
major motivation of this paper.

\section{methodology}

To define the QSL, the physical scenario and target needs to be clarified first.
Bloch sphere is a natural representation to show the geometry of quantum mechanics.
It is known that a $N$-dimensional density matrix $\rho$ can be expressed by a
Bloch vector via the equation below~\cite{Nielsen2000}
\begin{equation}
\rho=\frac{1}{N}\Big(\openone+\sqrt{\frac{N(N-1)}{2}}\vec{r}\cdot
\vec{\lambda}\Big),
\end{equation}
where $\vec{r}$ is the Bloch vector, $\openone$ is the identity matrix and $\vec{\lambda}$
is a $(N^{2}-1)$-dimensional vector of $\mathfrak{su}(N)$ generators. Through
this paper, the target we consider is defined via the angle~\cite{Campaioli2018}
\begin{equation}
\theta(t,\vec{r}) := \arccos \left(\frac{\vec{r}
\cdot \vec{r}(t)}{|\vec{r}||\vec{r}(t)|}\right),
\end{equation}
where $\vec{r}$ and $\vec{r}(t)$ are the initial and evolved states.
$\theta\in (0,\pi]$. The physical scenario for the QSL is~\emph{evolving some initial
state $\vec{r}$ with a Hamiltonian $H$ to any state satisfying the target angle $\Theta$}
($\Theta$ is a known fixed angle defined by above equation).

For a Hamiltonian $H$, it is possible that not all states in the state space
can fulfill the target, yet this fact was widely neglected in the previous
studies of the QSL based on inequalities. Here we first define a set $\mathcal{S}$
as the set of initial states that can fulfill the target angle, \emph{i.e.},
\begin{equation}
\mathcal{S}:=\{\vec{r}|\theta(t,\vec{r})=\Theta, \exists t\}.
\end{equation}
Similarly, we also define the set of reachable target states as
\begin{equation}
\mathcal{D}:=\{\vec{r}_{\mathrm{tar}}|\Theta=\arccos \left(\frac{\vec{r}\cdot
\vec{r}_{\mathrm{tar}}}{|\vec{r}||\vec{r}_{\mathrm{tar}}|}\right),\vec{r}\in\mathcal{S}\}.
\end{equation}
Here are some observations on $\mathcal{S}$ and $\mathcal{D}$.
\begin{prop}
$\mathcal{S}=\mathcal{D}$ for periodic evolutions. \label{prop:periodic_evolution}
\end{prop}
This can be easily proved since the dynamical trajectories of periodic evolution
are closed. Any two states on the same trajectory can evolve to each other.
\begin{prop} \label{prop:twoTheta}
For two target angles $\Theta_1,\Theta_2 \neq \pi$, if the dynamics of the quantum
states is continuous, then $S(\Theta_1)\subset S(\Theta_2)$ for $\Theta_1>\Theta_2$.
\end{prop}
In the case that the dynamics is continuous, the inner product between the initial
and evolved states is also continuous, therefore, if the state can reach the target
angle $\Theta$, it can also reach all the target angles smaller than $\Theta$.
One exception here is $\Theta=\pi$. In some open systems, it is possible that
some Bloch vectors only change the length. In this case, when the states evolve
through the zero vector and then change direction, it can still reach the
angle $\pi$, yet the inner production is not continuous during the evolution.

The time-independent Hamiltonian is one of the major subjects in the study of the QSL.
Here we provide an explicit expression of $\mathcal{S}$ for any dimensional
time-independent Hamiltonians under unitary evolution (the derivation is in
Appendix~\ref{sec:apx_S}).
\begin{prop} \label{prop:S_timeindependent}
For a $N$-dimensional time-independent Hamiltonian under unitary evolution,
one expression of $\mathcal{S}$ in the energy basis $\{|E_i\rangle\}$ is
\begin{eqnarray}
\mathcal{S} &=& \Bigg\{\vec{r}~\Big|1-\cos\Theta=\frac{1}{|\vec{r}|^2}\sum^{N-1}_{n=1}
\sum^{n-1}_{i=0}\left(1-\cos\left[(E_n-E_i)t\right]\right) \nonumber \\
& & \times \left(r^{2}_{n^{2}+2i-1}+r^{2}_{n^2+2i}\right), \exists t \Bigg\},
\end{eqnarray}
where $E_i$ (with corresponding eigenstate $|E_i\rangle$) is the $i$th energy
eigenvalue (we assume $E_i\leq E_j$ for $i\leq j$) and $r_i$ is the $i$th entry
of $\vec{r}$.
\end{prop}

%========================Figure==========================
\begin{figure}[tp]
\centering
\includegraphics[width=8cm]{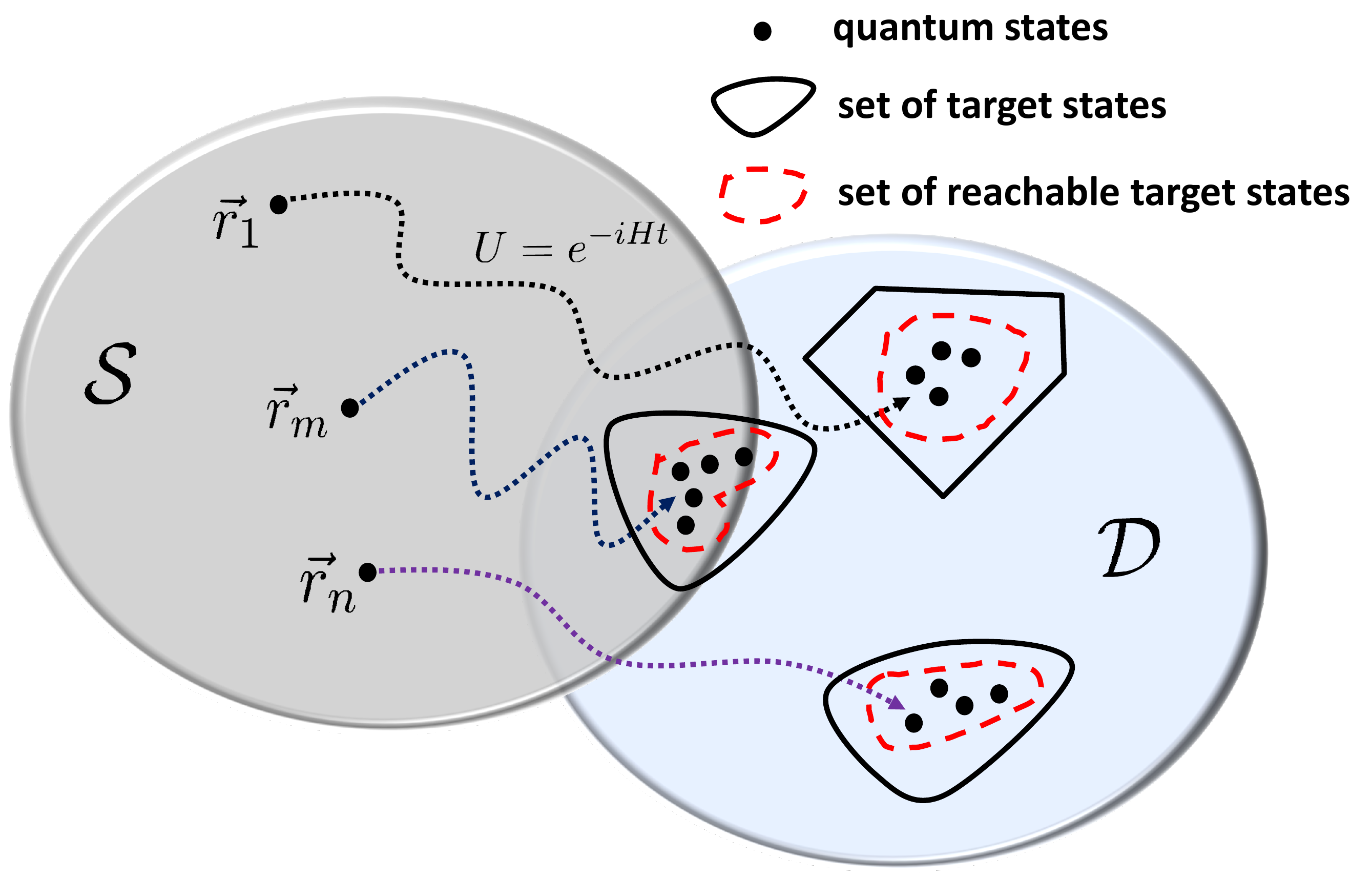}
\caption{(Color online) Schematic for operational definition of the QSL. For any state
in $\mathcal{S}$, there exists a subset of $D$ (area within the solid black line)
including all the target states (states satisfying the target angle $\Theta$),
and some of them (area within the dashed red line) are reachable for a specific $H$.
The minimum evolution time for all states in $S$ to reach the target states is
the operational QSL $\tau$.}
\label{Fig:S_and_D_set}
\end{figure}
%========================================================

With the assistance of $\mathcal{S}$, now we are in a position to introduce the
operational definition of the QSL.
\begin{definition}
The QSL $\tau$ is defined as the minimum evolution time to fulfill
$\Theta$ for any $\vec{r}\in\mathcal{S}$, i.e.,
\begin{eqnarray}
\tau &:=& \min_{\vec{r}\in\mathcal{S}} t \nonumber \\
& & \mathrm{s.t.}~\theta(t,\vec{r})=\Theta.
\end{eqnarray}
\end{definition}
This operational definition requires two steps to measure the QSL: (1) find the
regime of the set $\mathcal{S}$ and (2) find the minimum evolution time to reach
the target angle for states in $\mathcal{S}$, as shown in Fig.~\ref{Fig:S_and_D_set}.
This definition makes the QSL measurable in physics. For a specific quantum system,
we can first find the regime of $\mathcal{S}$ either theoretically or experimentally,
then experimentally prepare enough initial states in $\mathcal{S}$ and measure the
corresponding evolution time to reach the target angle. At last, the minimum time
of them is just the QSL we seek. This definition has two obvious advantages:
(i) it is guaranteed to be attainable by the definition and (ii) it is state-independent,
which means it only reflects the fundamental property of the Hamiltonian structure
and decoherence.

Another benefit with the assistance of $\mathcal{S}$ is that we can now define
a finite guaranteed time to reach the target angle as the maximum time in $\mathcal{S}$.
\begin{definition}
The guaranteed time to reach the target angle $\Theta$ is defined as
\begin{eqnarray}
\zeta &:=& \max_{\vec{r}\in\mathcal{S}} t \nonumber \\
& & \mathrm{s.t.}~\theta(t,\vec{r})=\Theta.
\end{eqnarray}
\end{definition}
It is impossible to define a finite guaranteed time without $\mathcal{S}$ in
general since the time for the states out of $\mathcal{S}$ to reach $\Theta$
is actually infinite. In the following we will discuss it in various scenarios,
including time-independent and time-dependent Hamiltonians and open systems.

\section{Time-independent Hamiltonians}

The first scenario we consider is time-independent Hamiltonians, for which we
have the following theorem.
\begin{theorem}
For a general multi-level system with a time-independent Hamiltonian $H$, the
operational definition of the QSL is
\begin{equation}
\tau =\frac{\Theta}{E_{\mathrm{max}}-E_{0}},
\end{equation}
where $E_{\mathrm{max}}$ and $E_{0}$ are the highest and lowest energies with
respect to $H$. This QSL $\tau$ can be attained by the states
\begin{equation}
\rho_{\mathrm{opt}}=\sum_{i}\frac{1}{N}|E_{i}\rangle\langle E_{i}|\!
+\xi|E_0\rangle\langle E_{\rm max}|\!+\xi^{*}|E_{\rm max}\rangle\langle E_0|
\label{eq:attainability}
\end{equation}
with the complex coefficient $\xi$ satisfying $|\xi|\in (0, 1/N]$.
\end{theorem}
The proof of this theorem based on Proposition~\ref{prop:S_timeindependent} is
given in Appendix~\ref{sec:apx_Nlevel}. For other states in $\mathcal{S}$ that
not in the form of Eq.~({\ref{eq:attainability}}), $\tau$ is a lower bound of
the corresponding evolution time to reach the target angle. In the following
we give several remarks on this theorem.

\textbf{Remark 1.} The attainable states are mixed states for $N\geq 3$. They
can only be pure in two-level systems by choosing $|\xi|=1/2$, which is the
reason why the bounds attainable for pure states, like MT and ML bounds,
can only be saturated in two-level systems~\cite{Deffner2017}.

\textbf{Remark 2.} It does not require a zero ground-state energy to be attainable.
In the case of two-level systems, the only case that $\tau_{\mathrm{MT}}$ and
$\tau_{\mathrm{ML}}$ are attainable, if the ground-state energy is set to be zero, then
$\tau=\min \tau_{\mathrm{MT}}=\min \tau_{\mathrm{ML}}$ for $\Theta=\pi$ (\emph{i.e.},
the orthogonal states as the target).

\textbf{Remark 3.} This bound can also be obtained by the bound
$\tau_{\mathrm{B}}=\Theta/Q$ ($Q$ is given in Eq.~(\ref{eq:tauB_Q}))~\cite{Campaioli2018}
with a proper choice of $\mathfrak{su}(N)$ generators and the optimization over
$\mathcal{S}$. The discussion is in Appendix~\ref{sec:apx_Nlevel}.

A corollary on the guaranteed time $\zeta$ can be immediately obtained for
periodic evolutions.
\begin{corollary} \label{corollary:zeta_periodic}
For time-independent Hamiltonians, the guaranteed time for a periodic evolution
with period $T$ is
\begin{equation}
\zeta=T-\tau.
\end{equation}
\end{corollary}

Define $\mathcal{S}^{(km)}$ ($m<k<N$) as a subset of $\mathcal{S}$ given in
Proposition~\ref{prop:S_timeindependent}, and all the legitimate states in
$\mathcal{S}^{(km)}$ satisfy $r^{2}_{n^{2}+2i-1}+r^{2}_{n^{2}+2i}$ is non-zero
for $i=m,n=k$ and zero for others subscripts, we have the following corollary.
\begin{corollary}
For all legitimate states in $\mathcal{S}^{(km)}$, the minimum time $\tau_{km}$
to reach target angle can be expressed by
\begin{equation}
\tau_{km}=\frac{\Theta}{E_k-E_m}.
\end{equation}
\end{corollary}

%========================Figure==========================
\begin{figure}[tp]
\centering
\includegraphics[width=8cm]{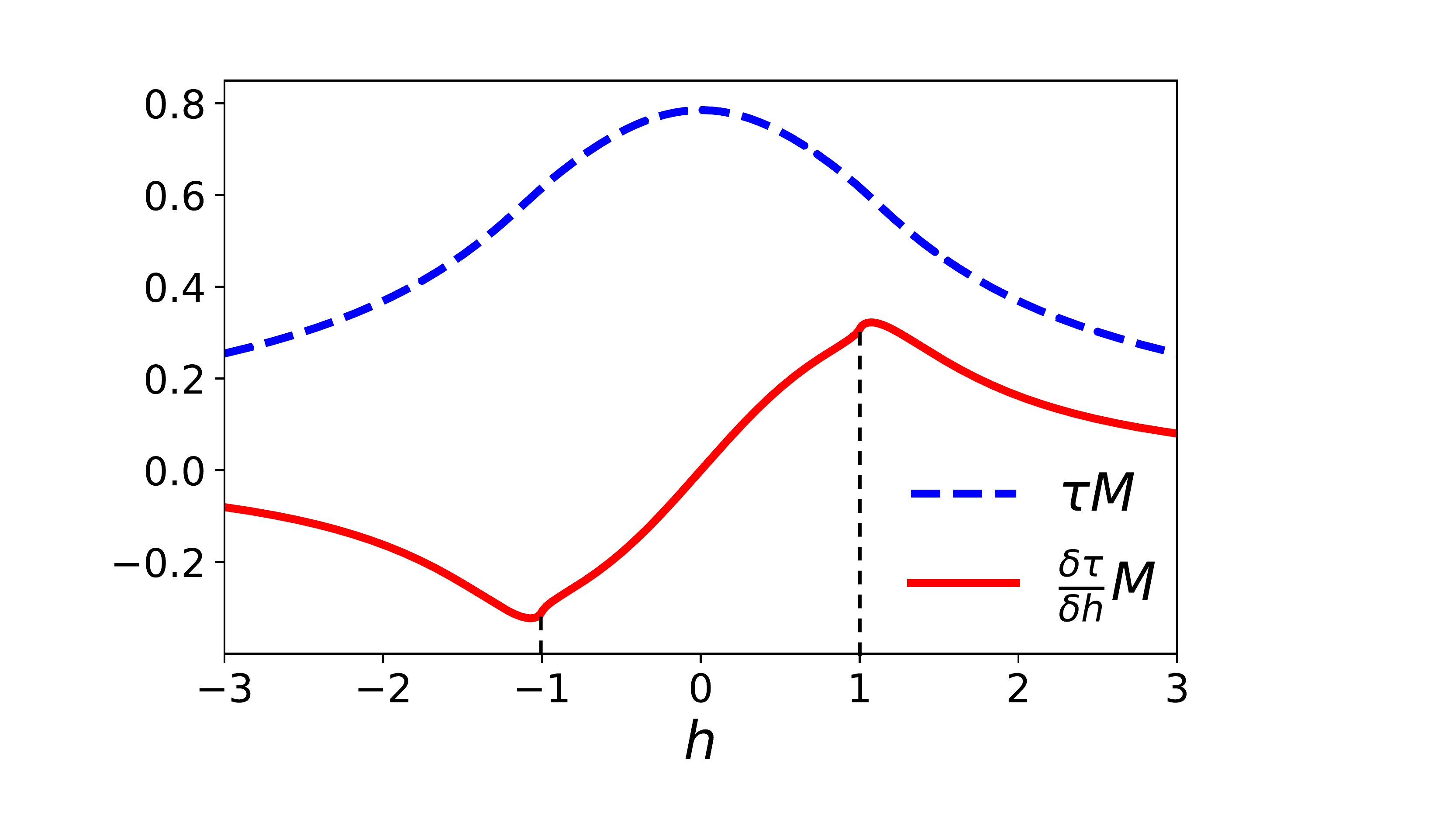}
\caption{(Color online) The QSL $\tau$ (dashed blue line) and its derivative with
respect to $h$ (solid red line) as functions of $h$ in one-dimensional transverse
Ising model. $J$ is set to be 1 in the plot. The target $\Theta=\pi/2$.}
\label{fig:Ising}
\end{figure}
%========================================================

Due to \textbf{Remark 2} that $\tau$ does not require a vanishing ground state
energy, many intriguing phenomena of the ground state can be exhibited in $\tau$,
such as the quantum phase transition~\cite{Heyl2017}. Here we use the one-dimensional
transverse Ising model as an example to show that the susceptibility of $\tau$ with
respect to the external field can be used as an indicator for the quantum phase transition.
The Hamiltonian of the model is
\begin{equation}
H = - J\left(\sum^{M}_{i=1}\sigma_{i}^{x}\sigma_{i+1}^{x}
+ h\sum^{M}_{i=1}\sigma_{i}^{z}\right),
\end{equation}
where $\sigma^{x}_{i}$ ($\sigma^{z}_{i}$) is the the Pauli X(Z) matrix for
the $i$th spin, $J$ is the interaction strength and $h= B/J$ with $B$ the
strength of the external field. $M$ is the spin number. Taking into account the
periodic boundary condition, the Hamiltonian above can be analytically solved
as $H/J=2 \sum_{k}\omega_{k} c_{k}^{\dagger}c_{k}
-\sum_{k}\omega_{k}$, where $\omega_{k}=\sqrt{1-2h\cos{k}+h^{2}}$,
and $c_{k}$ $(c^{\dagger}_{k})$ is a fermionic annihilation (creation) operator.
$k=2\pi n/M$ with $n=0,\pm 1,\cdots,\pm \frac{1}{2}(M-1)$ for odd $M$ and
$n=\pm \frac{1}{2},\pm \frac{3}{2},\cdots,\pm \frac{1}{2}(M-1)$ for even $M$.
The ground-state energy is $E_{0}/J=-\sum_{k}\omega_{k}$ and the highest
energy is $E_{\mathrm{max}}/J=\sum_{k}\omega_{k}$. At the thermodynamic
limit (details in Appendix~\ref{apx:Ising}), the QSL reads
\begin{equation}
\tau=\frac{\pi\mathrm{sgn}(1+h)\Theta/J}{4M(h+1)E\left(\frac{4h}{(h+1)^2}\right)},
\end{equation}
where $\mathrm{sgn}(\cdot)$ is the sign function and $E(\cdot)$ is the
complete elliptic function of the second kind. Furthermore, the susceptibility
of $\tau$ with respect to $h$ is
\begin{eqnarray}
\frac{\delta \tau}{\delta h} &=& \frac{\mathrm{sgn}(h+1)\pi \Theta/J}{8M h(h+1)^2 E^{2}
\!\left(\frac{4h}{(h+1)^2}\right)}\Bigg[\!(h+1)E\!\left(\frac{4h}{(h+1)^2}\right)
\nonumber \\
& & +(h-1)K\!\left(\frac{4h}{(h+1)^2}\right)\!\Bigg],
\end{eqnarray}
where $K(\cdot)$ is the complete elliptic function of the first kind.

The QSL and its susceptibility with respect to $h$ are shown in Fig.~\ref{fig:Ising},
in which the largest $\tau$ is always obtained at $h=0$.
More importantly, $\delta \tau/\delta h$ is not smooth at $h=\pm1$,
which is due to the well-known fact that $h=\pm 1$ are the critical points.
Thus, the susceptibility of the QSL is an observable to detect the phase transition.
The corresponding scheme is to prepare the system in the state
$\rho_{\mathrm{opt}}$ and then measure the change of the evolution time when the
target angle is reached. This scheme is robust to the dephasing noise during
the state preparation because $\tau$ can be attained by any reasonable nonzero
value of $\eta$.

\begin{figure}[tp]
\centering
\includegraphics[width=8cm]{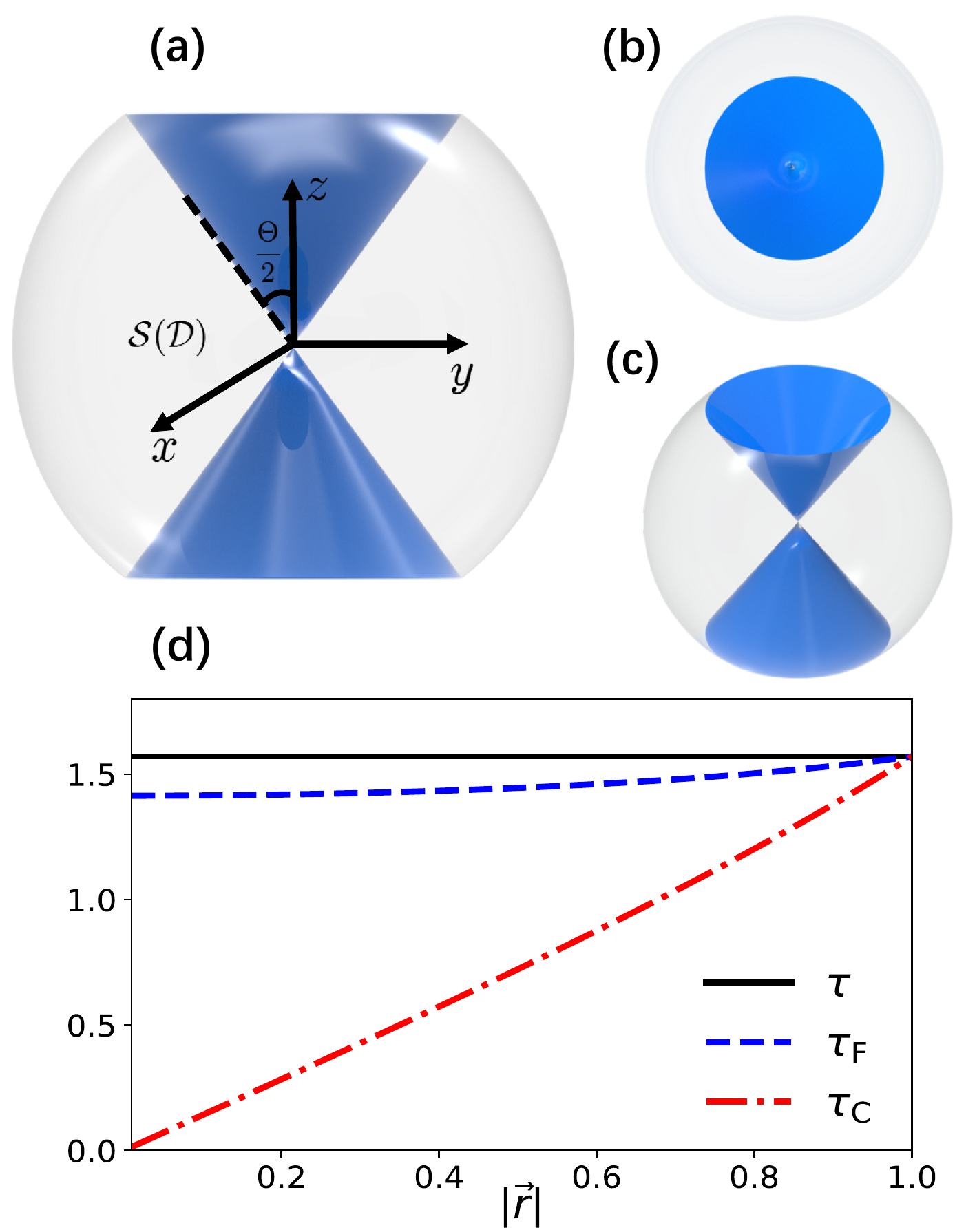}
\caption{(Color online) The set $\mathcal{S}(\mathcal{D})$ for a
two-level system in Bloch sphere from (a) front view; (b) top view;
(c) oblique view. The light gray area is $\mathcal{S}(\mathcal{D})$.
The states in the blue cones cannot fulfill the target angle $\Theta$.
(d) Comparison among $\tau$ (solid black line), $\tau_{\mathrm{F}}$
(dashed blue line) and $\tau_{\mathrm{C}}$ (dash-dotted red line) for
the initial states in the $xy$ plane. The target angle $\Theta=\pi/2$.}
\label{fig:TLS_S}
\end{figure}

Two-level systems are the earliest systems in the study of QSL and also the
only case in which $\tau_{\mathrm{C}}$ and $\tau_{\mathrm{F}}$ are attainable.
For two-level systems, any state can be expressed via the Bloch vector
\begin{equation}
\vec{r}(\eta,\alpha,\varphi)=\eta(\sin\alpha\cos\varphi,
\sin\alpha\sin\varphi,\cos\alpha),
\label{eq:blochvec_qubit}
\end{equation}
where $\eta\in[0,1]$, $\alpha\in[0,\pi]$ and $\varphi\in[0,2\pi]$.
Since the unitary evolution of a two-level system is periodic, $\mathcal{S}$ is
equivalent to $\mathcal{D}$ in this case according to Proposition~\ref{prop:periodic_evolution}.
Furthermore, we have the following corollary.
\begin{corollary} \label{corollary:qubit}
For a 2-dimensional time-independent Hamiltonian under unitary evolution,
the set $\mathcal{S}(\mathcal{D})$ in the Bloch representation is
\begin{equation}
\left\{\vec{r}(\eta,\alpha,\varphi)\Big|\eta\in (0,1],
\alpha\!\in\!\left[\frac{\Theta}{2},\pi-\frac{\Theta}{2}\right]\!\!,
\varphi\!\in\![0,2\pi]\right\}.
\label{eq:S_qubit}
\end{equation}
Let $E_{0}$ and $E_{1}$ be the ground and excited energies of the Hamiltonian,
then the operational definition of the QSL is
\begin{equation}
\tau = \frac{\Theta}{E_{1}-E_{0}}.
\label{eq:tau_2level}
\end{equation}
\end{corollary}
A thorough discussion of this case from a geometric perspective in Bloch sphere
is in Appendix~\ref{sec:axp_twolevel}. In the Bloch sphere (with $|E_{1}\rangle$
the north pole), $\mathcal{S}$ is the light gray area in Fig.~\ref{fig:TLS_S}(a).
All states in Bloch sphere apart from the double cone with the apex angle
$\Theta$ belong to $\mathcal{S}$. It can be seen that the volume of $\mathcal{S}$
shrinks with the increase of $\Theta$, which can be explained via Proposition~\ref{prop:twoTheta}.
Physically, most states in $\mathcal{S}$ here have two states on the dynamical
trajectory satisfying the target angle $\Theta\neq \pi$ and one for $\Theta=\pi$.

In regard to the QSL, $\tau$ can be attained by any state in the $xy$ plane
apart from the original point. A major difference between $\tau$
in Eq.~(\ref{eq:tau_2level}) and $\tau_{\mathrm{C}}$, $\tau_{\mathrm{F}}$ is that
$\tau$ is attainable for both pure and mixed states. Figure~\ref{fig:TLS_S}(d)
compares $\tau$ (solid black line), $\tau_{\mathrm{C}}$ (dash-dotted red line)
and $\tau_{\mathrm{F}}$ (dashed blue line) as a function of $|\vec{r}|$ for
the states in the $xy$ plane (in which they are all irrelevant to $\varphi$).
It shows $\tau$ is always the tightest bound for any value of $|\vec{r}|$ since
it is always attainable in this plane. When $|\vec{r}|=1$, both $\tau_{\mathrm{C}}$
and $\tau_{\mathrm{F}}$ coincide with $\tau$, confirming the fact that they are only
attainable for pure states in this case. Meanwhile, since the dynamics in this
case is periodic with the period $\frac{2\pi}{E_1-E_0}$, the guaranteed time
then reads
\begin{equation}
\zeta=\frac{2\pi-\Theta}{E_1-E_0}
\end{equation}
according to Corollary~\ref{corollary:zeta_periodic}.

%=======================Figure======================
%===================================================

\section{Time-dependent Hamiltonians}

%========================Figure==========================
\begin{figure*}[tp]
\centering
\includegraphics[width=16cm]{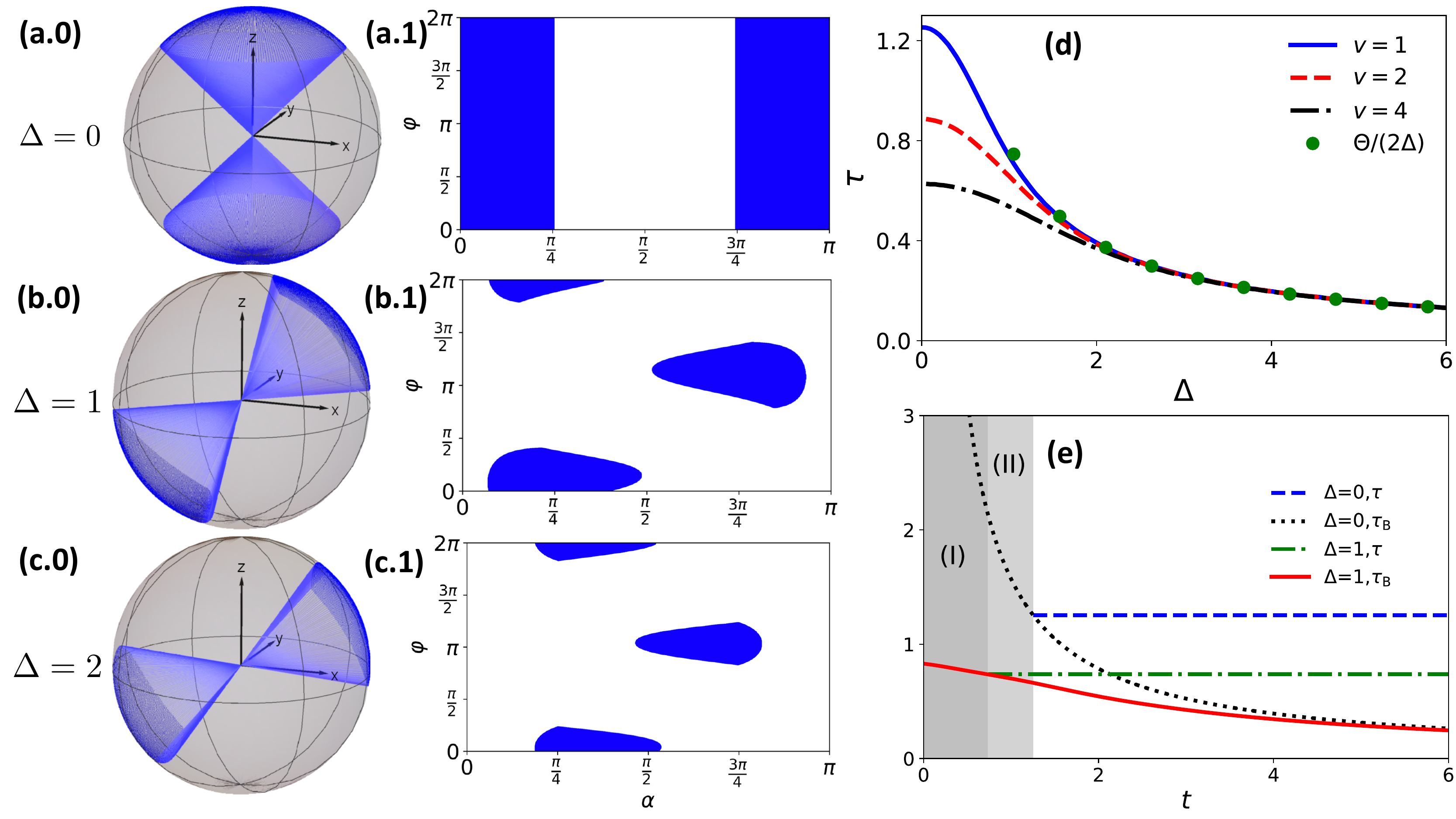}
\caption{(Color online) (a-c) The set $\mathcal{S}$ for (a) $\Delta=0$;
(b) $\Delta=1.0$ and (c) $\Delta=2.0$ in Landau-Zener model. (a.0), (b.0)
and (c.0) show the distributions of $\mathcal{S}$ in Bloch spheres.
(a.1), (b.1) and (c.1) show the distributions of $\mathcal{S}$ as a
function of $\alpha$ and $\varphi$. The white (and gray) and blue areas
are the regimes that the target angle $\Theta=\pi/2$ can and cannot 
fulfill, respectively. $v$ is set to be 1 in the plots.
(d) The QSL $\tau$ as a function of $\Delta$ for $v=1$ (solid blue line),
$v=2$ (dashed red line) and $v=4$ (dash-dotted black line) in Landau-Zener model.
The green dots represent the analytical solution of $\tau$ for large $\Delta$.
The target in the plot is $\Theta=\pi/2$. (e) Comparison between $\tau$ and
$\tau_{\mathrm{B}}$ for different values of $\Delta$. The dashed blue and dotted
black lines represent $\tau$ and $\tau_{\mathrm{B}}$ for $\Delta=0$
and the dash-dotted green and solid red lines represent $\tau$ and
$\tau_{\mathrm{B}}$ for $\Delta =1$. $v$ is set to be 1. The target
angle $\Theta=\pi/2$.}
\label{Fig:S_LZ}
\end{figure*}
%========================================================

Finding the QSL for time-dependent Hamiltonians is always a core task in the
studies of this field. In the previous researches, most theoretical tools for
time-dependent Hamiltonians are formal inequalities with respect to time and the
bounds contain a average process over the time, which make them time-dependent.
This is not reasonable as already discussed in the introduction. Here we will show
that the operational definition of the QSL does not have such problems and can reveal
the true physics behind the QSL. We take the Landau-Zener model as an example,
of which the Hamiltonian is
\begin{equation}
H=\Delta \sigma_x+vt\sigma_z,
\end{equation}
where $\Delta$ and $v$ are time-independent parameters. In the following we take
the eigenstate of the positive eigenvalue of $\sigma_z$ as the north pole of Bloch sphere.
In the case that $\Delta=0$, $\mathcal{S}$ is also in the form of Eq.~(\ref{eq:S_qubit})
since the dynamics is still the rotation around the $z$ axis, which is also numerically
confirmed in Fig.~\ref{Fig:S_LZ}(a.0).

For a non-vanishing $\Delta$, the analytical expression of $\mathcal{S}$ is hard
to obtain, therefore we provide the numerical results in Fig.~\ref{Fig:S_LZ}(b)
and (c) for $\Delta=1.0$ and $2.0$, respectively. Figure~\ref{Fig:S_LZ}(b.0)
and (c.0) show the distributions of $\mathcal{S}$ in Bloch spheres. For the sake of a
better presentation, we replot $\mathcal{S}$ as a function of $\alpha$ and $\varphi$,
defined in Eq.~(\ref{eq:blochvec_qubit}), in Fig.~\ref{Fig:S_LZ}(a.1), (b.1) and (c.1).
The distribution of $\mathcal{S}$ is not affected by $\eta$ since the dynamics
is unitary. The gray areas in Bloch spheres and white areas in $\alpha$-$\varphi$
plots represent the regimes of $\mathcal{S}$ and the blue areas are the set of
states that cannot reach the target angle ($\bar{\mathcal{S}}$). The target
angle $\Theta=\pi/2$ and $v=1$ in all plots. One may notice that $\bar{\mathcal{S}}$
is central symmetric about the original point, which is due to the fact that the
dynamical trajectories of a pair of central symmetric initial states are also
central symmetric (graphically shown in Appendix~\ref{apx:LZ}). The area of
$\bar{\mathcal{S}}$ shrinks with the increase of $\Delta$, indicating that a
larger $\Delta$ allows more states to reach the target angle in this case.

In the case of $\Delta=0$, the operational definition of the QSL can be analytically
obtained (details in Appendix~\ref{apx:LZ}) as follows
\begin{equation}
\tau=\sqrt{\frac{\Theta}{v}},
\end{equation}
which is only the function of Hamiltonian parameters $v$ and the target angle,
rather than the function of time. This result confirms our argument that the QSL
for time-dependent Hamiltonians should not be a function of time. $\tau$ can be
attained by any state in the $xy$ plane apart from the original point. Furthermore,
since the dynamics here is still periodic with the period $T=\sqrt{2\pi/v}$,
the guaranteed time $\zeta$ is
\begin{equation}
\zeta = \frac{1}{\sqrt{v}}\left(\sqrt{2\pi}-\sqrt{\Theta}\right).
\end{equation}

The operational definition of the QSL for a non-vanishing $\Delta$ is numerically
calculated and shown in Fig.~\ref{Fig:S_LZ}(d) as a function of $\Delta$ for different
values of $v$. One can see $\tau$ always decays with the increase of $\Delta$ and $v$.
For a large $\Delta$, $\tau$ is independent of $v$, which is due to the fact that
in this regime $\Delta\sigma_x$ is the dominant term in Hamiltonian and the QSL
reduces to $\Theta/(2\Delta)$ (green dots in Fig.~\ref{Fig:S_LZ}(d)) according to
Corollary~\ref{corollary:qubit}.

In the meantime, $\tau_{\mathrm{B}}$ in this case can be calculated as
\begin{equation}
\tau_{\mathrm{B}} = \frac{\Theta}{vt}\sqrt{\frac{|\vec{r}|^2}{|\vec{r}|^2-r^2_z}},
\end{equation}
which is inverse propositional to the time $t$. $r_z$ is the third entry of the Bloch vector.
For the states in the $xy$ plane where $\tau$ is attainable, $\tau_{\mathrm{B}}=\Theta/(vt)$
is still related to the time. Figure~\ref{Fig:S_LZ}(e) compares the performances
of $\tau$ and $\tau_{\mathrm{B}}$ for different values of $\Delta$. The dashed-blue
and dash-dotted green lines represent $\tau$ for $\Delta=0$ and $1$, respectively.
And the dotted black and solid red lines represent $\tau_\mathrm{B}$ for $\Delta=0$
and $1$. The initial states of $\tau_\mathrm{B}$ are taken as those that can reach $\tau$.
The target angle $\Theta=\pi/2$ and $v$ is set to be $1$. From this figure,
one can see that after the time $\tau$, $\tau$ is always tighter than $\tau_\mathrm{B}$
since $\tau$ is true and attainable. In the case of $\Delta=0$, the target
angle $\Theta$ cannot be fulfilled by the evolution time in the gray regimes (I)
and (II), which means the evolution time to reach $\Theta$ in this regime is
actually \emph{infinity} in mathematics. Therefore, any finite value can provide
a mathematically correct bound in this case, as given by $\tau_\mathrm{B}$ and other
bounds based on the same philosophy (similar things happen in the regime (I) for
$\Delta=1$). However, these bounds themselves cannot provide this information and
sometimes may mislead the true physics behind the mathematics.

\section{Open systems}

The QSL in open systems is intriguing yet more complicated compared to the
unitary evolution. Many works attempted to provide attainable bounds for open
systems. In regard to the Bloch representation, Campaioli \emph{et al.}~\cite{Campaioli2019}
used the distance between two Bloch vectors to derive a bound of the QSL. Here we
show the performance of the operational definition of the QSL in open systems.

A large number of quantum dynamics of open systems is governed by the following
master equation
\begin{equation}
\partial_{t}\rho\left(t\right)\!=\!-i\left[H,\rho\right]\!+\!\sum_i\!\gamma_{i}
\!\left[L_i\rho(t)L^{\dagger}_i-\frac{1}{2}\{L^{\dagger}_i L_i,\rho(t)\}\right],
\end{equation}
where $L_i$ is the $i$th Lindblad operator depicting certain decay mode. For a
time-independent Hamiltonian under Markovian dynamics, \emph{i.e.}, $\gamma_i$
is time-independent for any subscript $i$, the dynamics of the corresponding
Bloch vector is an affine map
\begin{equation}
\vec{r}(t)=e^{\mathcal{M}^{\mathrm{T}}t}(\vec{r}-\vec{l}~)+\vec{l},
\end{equation}
where $\mathcal{M}$ and $\vec{l}$ are real and the specific expressions are given in
Appendix~\ref{apx:sec_opensys}. For this dynamics, the set $\mathcal{S}$ is of the form
\begin{equation}
\mathcal{S}=\left\{\vec{r}\,\Big|\cos\Theta=\frac{\vec{r}^{\,\mathrm{T}}
e^{\mathcal{M}^{\mathrm{T}}t}(\vec{r}-\vec{l}~)+\vec{r}^{\,\mathrm{T}}\vec{l}}
{\big|e^{\mathcal{M}^{\mathrm{T}}t}(\vec{r}-\vec{l}~)+\vec{l}\,\big||\vec{r}|},
\exists t\right\}.
\end{equation}

The first example we consider is the following master equation
\begin{eqnarray}
\partial_{t}\rho &=&-i\left[H,\rho\right]+\gamma_{+} \left[\sigma_{+}\rho\sigma_{-}
-\frac{1}{2}\{\sigma_{-}\sigma_{+},\rho\}\right] \nonumber \\
& & +\gamma_{-} \left[\sigma_{-}\rho\sigma_{+}-\frac{1}{2}\{\sigma_{+}\sigma_{-},\rho\}\right],
\label{eq:sigma_deco}
\end{eqnarray}
where the Hamiltonian $H=\frac{1}{2}\omega_0\sigma_z$ with $\sigma_{x,y,z}$ a
Pauli matrix, and $\sigma_{\pm}=\frac{1}{2}(\sigma_x\pm i\sigma_y)$. This model
can depict some important physical processes like the spontaneous emission
($\gamma_+=0$) and finite-temperature thermodynamics. Our first concern in open
systems is how $\mathcal{S}$ is affected by the decoherence. For the dynamics
governed by Eq.~(\ref{eq:sigma_deco}), $\mathcal{S}$ is of the form
\begin{equation}
\left\{\!\vec{r}(\eta,\alpha,\varphi)\!\big|\!\cos\Theta
\!=\!\frac{\sin^{2}\alpha\cos\left(\omega_{0}t\right)\!+\chi\cos\alpha}
{\sqrt{\sin^{2}\alpha+\chi^2}}, \exists t\!\right\}\!\!, \label{eq:sigma_S}
\end{equation}
where $\chi=e^{-\frac{1}{2}\gamma_{\mathrm{f}}t} \cos\alpha+\frac{2\gamma_{\mathrm{d}}}
{\eta \gamma_{\mathrm{f}}}\sinh\left(\frac{1}{2}\gamma_{\mathrm{f}}t\right)$
with $\gamma_{\mathrm{f}}=\gamma_{+}+\gamma_{-}$ and $\gamma_{\mathrm{d}}
=\gamma_{+}-\gamma_{-}$. The details of the calculation are given in Appendix~\ref{apx:sec_opensys}.
Notice that the constrain in Eq.~(\ref{eq:sigma_S}) does not involve $\varphi$,
which means in the Bloch sphere $\mathcal{S}$ is axial symmetric around the $z$ axis.

\begin{figure}[tp]
\centering
\includegraphics[width=8cm]{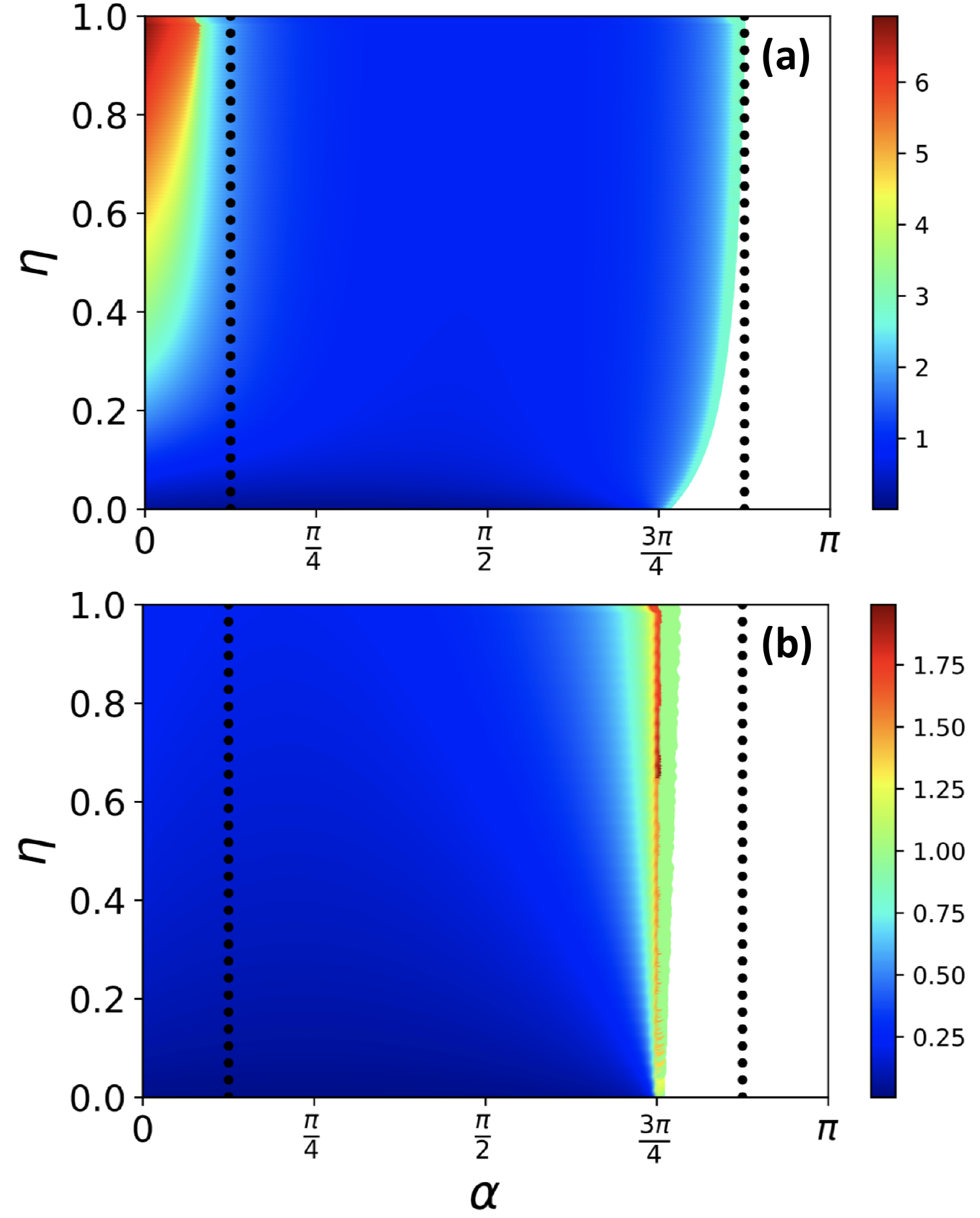}
\caption{(Color online) The set $\mathcal{S}$ and the evolution
time to reach the target angle as a function of $\alpha$ and
$\eta$ for (a) Markovian and (b) non-Markovian dynamics in the
case of spontaneous emission. The colored areas represent $\mathcal{S}$
and the values are corresponding minimum evolution time to reach
the target angle. The regime between the dashed black lines is
$\mathcal{S}$ for unitary evolution. $\omega_0$ is set to be $1$
and $\Theta=\pi/4$ in all plots. }
\label{fig:S_spon_emission}
\end{figure}

In the case of spontaneous emission ($\gamma_{+}=0$, $\gamma_{-}=\gamma$), the
distribution of $\mathcal{S}$ (colored area) and the corresponding values of the 
minimum time to reach the target angle $\Theta$ are given in
Fig.~\ref{fig:S_spon_emission}(a) as a function of $\alpha$ and $\eta$.
$\Theta=\pi/4$ and $\gamma=0.1$ in this plot. The area between the dotted black
lines is $\mathcal{S}$ under unitary evolution. It can be seen that the area of
$\mathcal{S}$ changes under the spontaneous emission. Affected by this decoherence,
some states with large $\alpha$ and small $\eta$ cannot reach the target angle anymore.
However, the beneficial part is that the states with small $\alpha$ can reach the
target angle now.

A more interesting phenomenon here is that the minimum evolution time reduces
with the decrease of $\eta$, which indicates that a lousy purity may speedup
the evolution to reach the target angle. To clarify the behavior of the QSL with
small $\eta$, we calculated corresponding $\tau$ analytically. For an acute
target angle, the operational definition of the QSL in this case approximates to
\begin{equation}
\tau \approx \frac{\delta\eta}{\gamma}\sin\Theta,
\end{equation}
where $\delta\eta$ is a small purity. The details of the calculation are in
Appendix~\ref{apx:sec_opensys}. $\tau$ in the above equation can be attained by
the states with $\alpha=\frac{\pi}{2}-\Theta$. In the studies of quantum information,
purity is always treated as a resource for many quantum information processings,
and the decoherence jeopardizes the purity and is harmful for those processings.
However, here our calculation shows that with respect to the QSL, the states with a
lousy purity may provide a shorter evolution time for the fulfillment of an acute
target angle, which is very counter-intuitive and has not been discovered by
other tools to the best of our knowledge. In this case, the increase of strength
may slightly reduce the size of $\mathcal{S}$ yet significantly enhances the
reduction of $\tau$.

%========================Figure==========================
%========================================================

The behaviors of the QSL with non-Markovian dynamics have drawn some attention in
recent years~\cite{Deffner2013,Sun2015,Meng2015,Mirkin2019}. The model of spontaneous
emission can also reveal the non-Markovian dynamics of damped Jaynes-Cummings
models, in which $\gamma=\gamma(t)$ is a time-dependent decay rate. In 2013,
Deffner and Lutz~\cite{Deffner2013} provided a very useful formula of the QSL for
purely initial states, and discussed the corresponding behavior in this case.
Here we also use it to show the performance of operational definition
of the QSL for non-Markovian dynamics. The only difference between non-Markovian and
Markovian dynamics in this model is that the decay rate $\gamma=\gamma(t)$ is
time-dependent. For the non-Markovian dynamics, $\mathcal{S}$ reads
\begin{equation}
\left\{\vec{r}\Big|\!\cos\Theta\!=\!
\frac{\sin^2\alpha\cos\left(\frac{1}{2}\mathrm{Im}(\Gamma)+\omega_0 t\right)
+\chi_1\cos\alpha}{\sqrt{\sin^2\alpha+f_1^2}}\right\}\!,
\end{equation}
where $\Gamma = \int^t_0\gamma(t_1)\mathrm{d}t_1$ and
$\chi_1=e^{-\frac{1}{2}\mathrm{Re}(\Gamma)}\cos\alpha
-\frac{2}{\eta}\sinh\left(\frac{1}{2}\mathrm{Re}(\Gamma)\right)$.
$\mathrm{Re}(\cdot)$ and $\mathrm{Im}(\cdot)$ represent the real and imaginary parts.
Figure~\ref{fig:S_spon_emission}(b) shows the distribution of $\mathcal{S}$ of
this non-Markovian dynamics. Compared to the Markovian dynamics, the area of
$\mathcal{S}$ shrinks and a state with $\alpha>3\pi/4$ can barely
reach the target angle. For the states with a small $\alpha$ and large $\eta$,
the minimum times to reach the target angle significantly reduce which means
non-Markovian dynamics can speedup the evolution to reach the target angle
for this parameter regime. A similar phenomenon that poor purity may
benefit the QSL is also observed here. Utilizing the similar calculation procedure
(details in Appendix~\ref{apx:sec_opensys}) in Markovian dynamics, $\tau$ satisfies
the following equation
\begin{equation}
\left(1-\frac{\lambda}{d}\right)\!e^{-\frac{1}{2}(d+\lambda)\tau}
\!+\!\left(1+\frac{\lambda}{d}\right)\!e^{\frac{1}{2}(d-\lambda)\tau}
=2e^{-\frac{1}{8}\delta\eta\sin\Theta},
\end{equation}
which is also attained by the states with $\alpha=\frac{\pi}{2}-\Theta$.
In this equation, $\tau$ monotonically reduces with the decrease of $\delta\eta$,
which means a small purity can indeed speed up the evolution to reach the target
angle in this non-Markovian dynamics.

%========================Figure==========================
\begin{figure}[tp]
\centering
\includegraphics[width=8cm]{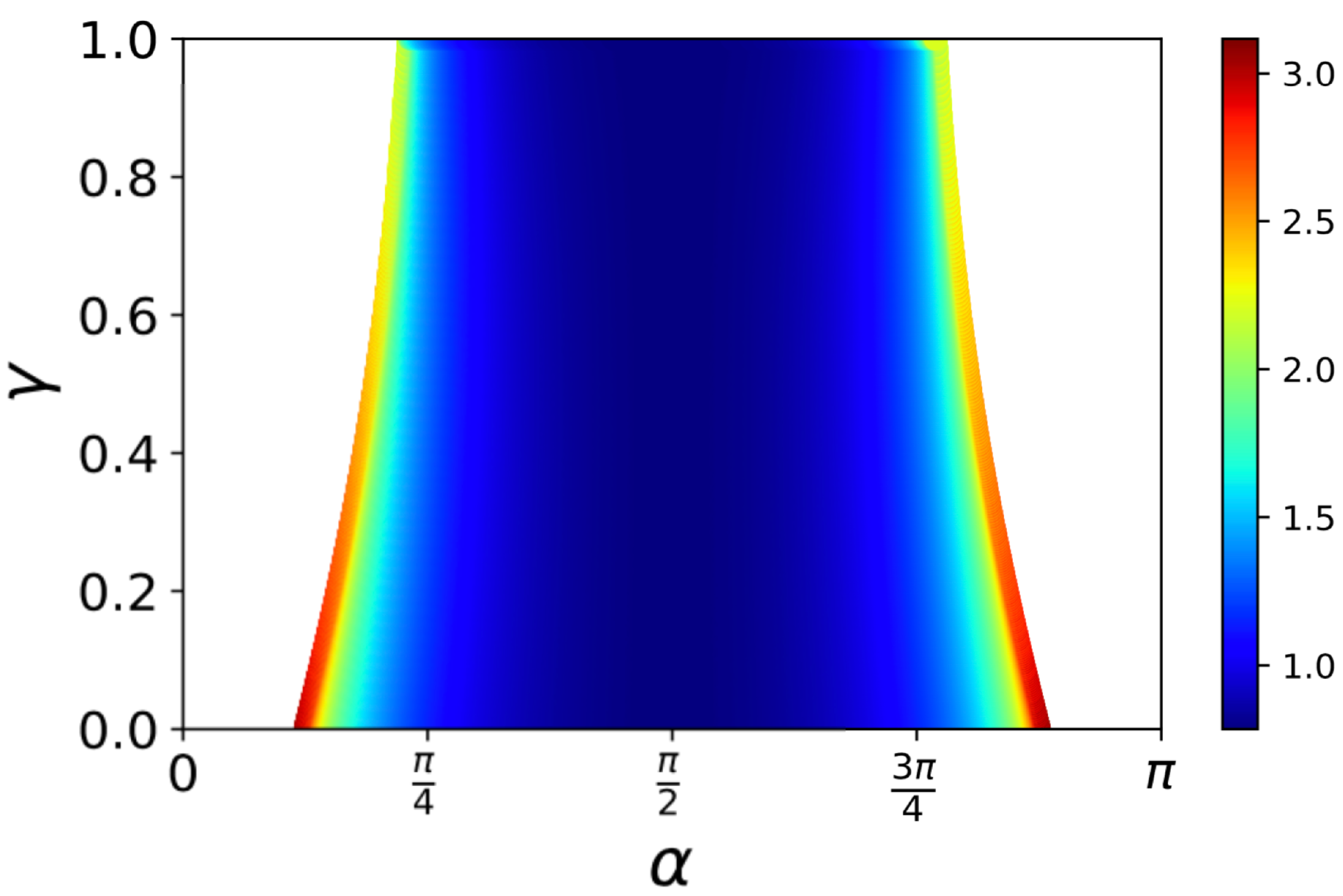}
\caption{(Color online) $\mathcal{S}$ and the minimum evolution
time to reach the target angle as a function of $\alpha$ and
decay rate $\gamma$ for $\Theta=\pi/4$ in the case of dephasing.
The colored areas represent $\mathcal{S}$ and the values are
corresponding minimum evolution time to reach the target angle.
$\omega_0$ is set to be 1. }
\label{fig:S_dephasing}
\end{figure}
%========================================================

Another example is the parallel dephasing
\begin{equation}
\partial_{t}\rho=-i\left[H,\rho\right]+\frac{\gamma}{2}\left(\sigma_{z}\rho
\sigma_{z}-\rho \right), \label{eq:dephasing}
\end{equation}
where $H$ is the same as that in the spontaneous emission. Dephasing is the
dominant decay mode for some physical processes like the recently discovered
collective phonons bundle emission~\cite{Bin2020}. In this dynamics,
$\mathcal{S}$ can be expressed by
\begin{equation}
\left\{\!\vec{r}(\eta,\alpha,\varphi)\Big|\!\cos\Theta\!=\!
\frac{1\!-\!\left[1\!-\!e^{-\gamma t}\cos(\omega_{0}t)\right]\!\sin^{2}\alpha}
{\sqrt{1-(1-e^{-2 \gamma t})\sin^{2}\alpha}}, \exists t \!\right\}\!.
\end{equation}
The details of the calculation are in Appendix~\ref{apx:sec_opensys}. Here only
$\alpha$ affects the distribution of $\mathcal{S}$, which means $\mathcal{S}$
always consists of two cones similar to the unitary evolutions. For example,
the distribution of $\mathcal{S}$ for $\Theta=\pi/4$ is given in Fig.~\ref{fig:S_dephasing}
as a function of $\alpha$ and $\gamma$, which shows that in this case the growth
of the decay rate will make $\mathcal{S}$ shrink, and the boundary of $\alpha$
moves towards $\pi/2$. In the case of $\Theta=\pi/2$, $\mathcal{S}$ reduces to
\begin{equation}
\left\{\vec{r}(\eta,\alpha,\varphi)|\eta\in(0,1],\alpha\in\left[\alpha_{\mathrm{c}},
\pi-\alpha_{\mathrm{c}}\right],\varphi\in[0,2\pi]\right\}.
\end{equation}
where the boundary $\alpha_{\mathrm{c}}=\arcsin\left(\frac{1}
{\sqrt{1+e^{-\gamma\pi/\omega_0}}}\right)$. In the case $\Theta=\pi$, $\mathcal{S}$
consists of all states in the $xy$ plane apart from the original point. It is
easy to see that $\mathcal{S}$ is not affected by the dephasing in this case.
For a reasonable value of $\gamma$, the operational definition of the QSL for
Eq.~(\ref{eq:dephasing}) reads $\tau=\Theta/\omega_0$, which can be
attained by all states with $\alpha=\pi/2$ and $\eta\neq 0$. This result coincides
with the unitary counterpart when $\omega_0$ represents the energy difference between
the excited and ground states, indicating that $\tau$ is not affected by the
parallel dephasing for a not extremely strong decay rate.

\section{Summary}

In conclusion, we have introduced an operational approach to the notion of the 
QSLs, which is state-independent and guaranteed to be attainable. With this
approach, we also define the guaranteed time for the fulfillment of the
target angle. The performances of this operational definition $\tau$ have been
thoroughly investigated in several scenarios. For time-independent Hamiltonians
under unitary evolutions, $\tau$ is inverse-proportional to the difference
between the highest and lowest energies. One advantage of this result is that
its attainability does not require a zero ground-state energy. The ground-state
energy contains fruitful phenomena in quantum physics like the quantum phase
transition. Therefore, the susceptibility of $\tau$ can be used as an indicator
of the quantum phase transition, which is demonstrated with the one-dimensional
transverse Ising model in the paper.

For the time-dependent Hamiltonians, the existing bounds of the QSL are basically
all related to the time, which is not reasonable in physics. We use the
Landau-Zener model as an example to show the true physics behind the QSL.
The analytical expression of $\tau$ is given for $\Delta=0$. With the
increase of $\Delta$, the value of $\tau$ approaches $\Theta/(2\Delta)$,
which is exactly the QSL for the time-independent term in the Hamiltonian.
The results in this case vividly clarify the fact that the QSL for
non-controlled time-dependent Hamiltonians should be irrelevant to the
evolution time.

The open systems is another important scenario for the research of the QSL.
The numerical and analytical calculations of $\tau$ in the case of the
spontaneous emission show a very interesting and counterintuitive phenomenon
that a lousy purity can benefit the reduction of the QSL, which, to the best
of our knowledge, has not been discovered with the existing tools. Furthermore,
this phenomenon occurs in both Markovian and non-Markovian dynamics, however,
the specific relations between $\tau$ and the purity are not exactly the same.

Different from conventional concerns about the QSL that focus on valid mathematical
tools, our operational approach emphasizes the physics behind the QSL, which
may provide the community another perspective for the study of fast dynamical
behaviors in quantum mechanics in the future. Moreover, the phenomena discovered
here would encourage experimenters to verify with many quantum systems. Finally, 
our approach should find broad applications in quantum technologies, such as
quantum control and parameter estimation in quantum metrology, and it should
carry over to arbitrary settings, including classical dynamics and stochastic
thermodynamics.

\begin{acknowledgments}
Y.S. and B.L. contributed equally to this work. The authors would like to thank
Prof. Adolfo del Campo for many insightful suggestions and help on improving
the introduction. We also thank Prof. Libin Fu, Prof. Xiao-Ming Lu,
Prof. Christiane Koch, Dr. Zibo Miao and Ms. Jinfeng Qin for helpful discussions.
This work was supported by National Natural Science Foundation of China through
Grant No.~11805073.
\end{acknowledgments}

\appendix

\section{The set $\mathcal{S}$ for time-independent Hamiltonians}
\label{sec:apx_S}

It is known that a $N$-dimensional density matrix can be expressed via the Bloch
vector as below
\begin{equation}
\rho=\frac{1}{N}\left(\openone+\sqrt{\frac{N(N-1)}{2}}\vec{r}\cdot\vec{\lambda}\right),
\end{equation}
where $\openone$ is the identity matrix, $\vec{r}$ is the Bloch vector satisfying
$|\vec{r}|\leq 1$, and $\vec{\lambda}$ is the vector of $\mathfrak{su}(N)$ generators.
For the unitary evolution, the evolved state $\rho(t)$ is
\begin{eqnarray}
\rho(t) &=& e^{-iHt}\rho e^{iHt} \nonumber \\
&=&\frac{1}{N}\left(\openone+\sqrt{\frac{N(N-1)}{2}}\vec{r}\cdot e^{-iHt}
\vec{\lambda}e^{iHt}\right).
\end{eqnarray}

Based on the property of $\mathfrak{su}(N)$ algebra, the unitary evolution of any
$\mathfrak{su}(N)$ generator can be expressed as the linear combination of all generators,
\emph{i.e.}, $e^{-iHt}\lambda_{i}e^{iHt}=\sum_{j}C_{ij}(t)\lambda_{j}$, which means
$\vec{r}\cdot e^{-iHt}\vec{\lambda}e^{iHt}=\sum_{ij}r_{i}C_{ij}(t)\lambda_{j}$.
This equation immediately leads to
\begin{equation}
\vec{r}(t)=C^{\mathrm{T}}(t)\vec{r},
\end{equation}
which is known as an unital affine map. $C_{ij}(t)$ can be further solved as
\begin{equation}
C_{ij}(t)=\frac{1}{2}\mathrm{Tr}\left(e^{-iHt}\lambda_{i}e^{iHt}\lambda_{j}\right),
\end{equation}
where the equation $\mathrm{Tr}(\lambda_i \lambda_j)=2\delta_{ij}$ has been used.
In the energy basis, $C_{ij}(t)$ reduces to
\begin{equation}
C_{ij}(t)=\frac{1}{2}\sum_{mk}e^{i(E_m-E_k)t}[\lambda_{i}]^{*}_{mk}[\lambda_{j}]_{mk}
\label{eq:apx_Ct}
\end{equation}
with $[\lambda_{j}]_{mk}$ the $mk$-th entry of $\lambda_{j}$ in the energy basis.
$E_i$ is the $i$th energy eigenvalue. In the following we use the specific energy
basis $\{|E_0\rangle,|E_1\rangle,\cdots,|E_{N-1}\rangle\}$, where we set
$E_0<E_1<\cdots<E_{N-1}$. In this basis with an appropriate representation of
$\mathfrak{su}(N)$ generators, the matrix $C(t)$ can be always expressed by
\begin{equation}
C(t)=\bigoplus^{N-1}_{n=1}V(n,t),
\label{eq:apx_Ct_matrix}
\end{equation}
where $V(n,t)=\left[\bigoplus^{n-1}_{i=0}M(E_n-E_i,t)\right]\oplus 1$ with
\begin{equation}
M(x,t)=\left(\begin{array}{cc}
\cos(xt) & -\sin(xt)\\
\sin(xt) & \cos(xt)
\end{array}\right).
\end{equation}
For example, for a two-level system, using the Pauli matrices as
the generators, $C(t)$ reads
\begin{equation}
C(t)=\left(\begin{array}{cc}
M(E_1-E_0,t) & 0 \\
0 & 1
\end{array}\right).
\end{equation}
For three-level systems, using the Gell-Mann matrices as the generators, $C(t)$
is of the form.
\begin{equation*}
\left(\begin{array}{ccccc}
M(E_1-E_0,t) & 0 & 0 & 0 & 0 \\
0 & 1 & 0 & 0 & 0 \\
0 & 0 & M(E_2-E_0,t) & 0 & 0 \\
0 & 0 & 0 & M(E_2-E_1,t) & 0 \\
0 & 0 & 0 & 0 & 1
\end{array}\right)\!\!.
\end{equation*}
The specific form of $\mathfrak{su}(4)$ generators with respect to
Eq.~(\ref{eq:apx_Ct_matrix}) is
\begin{eqnarray}
\lambda_0 &=& \left(\begin{array}{cccc}
0 & 1 & 0 & 0 \\
1 & 0 & 0 & 0 \\
0 & 0 & 0 & 0 \\
0 & 0 & 0 & 0
\end{array}\right)\!\!,
\lambda_1 = \left(\begin{array}{cccc}
0 & -i & 0 & 0 \\
i & 0 & 0 & 0 \\
0 & 0 & 0 & 0 \\
0 & 0 & 0 & 0
\end{array}\right)\!\!, \nonumber \\
\lambda_2 &=& \left(\begin{array}{cccc}
1 & 0 & 0 & 0 \\
0 & -1 & 0 & 0 \\
0 & 0 & 0 & 0 \\
0 & 0 & 0 & 0
\end{array}\right)\!\!,
\lambda_3 = \left(\begin{array}{cccc}
0 & 0 & 1 & 0 \\
0 & 0 & 0 & 0 \\
1 & 0 & 0 & 0 \\
0 & 0 & 0 & 0
\end{array}\right)\!\!,
\end{eqnarray}
and
\begin{eqnarray}
\lambda_4 &=& \left(\begin{array}{cccc}
0 & 0 & -i & 0 \\
0 & 0 & 0 & 0 \\
i & 0 & 0 & 0 \\
0 & 0 & 0 & 0
\end{array}\right)\!\!,
\lambda_5 = \left(\begin{array}{cccc}
0 & 0 & 0 & 0 \\
0 & 0 & 1 & 0 \\
0 & 1 & 0 & 0 \\
0 & 0 & 0 & 0
\end{array}\right)\!\!, \nonumber \\
\lambda_6 &=& \left(\begin{array}{cccc}
0 & 0 & 0 & 0 \\
0 & 0 & -i & 0 \\
0 & i & 0 & 0 \\
0 & 0 & 0 & 0
\end{array}\right)\!\!,
\lambda_7 \!=\! \frac{1}{\sqrt{3}}\!\!\left(\begin{array}{cccc}
1 & 0 & 0 & 0 \\
0 & 1 & 0 & 0 \\
0 & 0 & -2 & 0 \\
0 & 0 & 0 & 0
\end{array}\right)\!\!\!,
\end{eqnarray}
and
\begin{eqnarray}
\lambda_8 &=& \left(\begin{array}{cccc}
0 & 0 & 0 & 1 \\
0 & 0 & 0 & 0 \\
0 & 0 & 0 & 0 \\
1 & 0 & 0 & 0
\end{array}\right)\!\!,
\lambda_9 = \left(\begin{array}{cccc}
0 & 0 & 0 & -i \\
0 & 0 & 0 & 0 \\
0 & 0 & 0 & 0 \\
i & 0 & 0 & 0
\end{array}\right)\!\!, \nonumber \\
\lambda_{10} &=& \left(\begin{array}{cccc}
0 & 0 & 0 & 0 \\
0 & 0 & 0 & 1 \\
0 & 0 & 0 & 0 \\
0 & 1 & 0 & 0
\end{array}\right)\!\!,
\lambda_{11} = \left(\begin{array}{cccc}
0 & 0 & 0 & 0 \\
0 & 0 & 0 & -i \\
0 & 0 & 0 & 0 \\
0 & i & 0 & 0
\end{array}\right)\!\!,
\end{eqnarray}
and
\begin{eqnarray}
\lambda_{12} &=& \left(\begin{array}{cccc}
0 & 0 & 0 & 0 \\
0 & 0 & 0 & 0 \\
0 & 0 & 0 & 1 \\
0 & 0 & 1 & 0
\end{array}\right)\!\!,
\lambda_{13} = \left(\begin{array}{cccc}
0 & 0 & 0 & 0 \\
0 & 0 & 0 & 0 \\
0 & 0 & 0 & -i \\
0 & 0 & i & 0
\end{array}\right)\!\!,
\end{eqnarray}
and $\lambda_{14}=\frac{1}{\sqrt{6}}\mathrm{diag}(1,1,1-3)$. For higher
dimension, the generators can be constructed similarly.

The period $T$ of the evolution is determined by the period of $C(t)$,
which requires that all the energy gaps are commensurable with each other.
For the case that $E_{i+1}-E_{i}$ is a constant $d$ for any $i$, the period
of $C(t)$ is $T=2\pi/d$.

Recalling that $\vec{r}(t)=C^{\mathrm{T}}(t)\vec{r}$, the angle between the
initial and evolved Bloch vectors is
\begin{equation}
\cos\theta = \frac{\vec{r}(t)\cdot\vec{r}}{|\vec{r}|^{2}}
=\frac{\vec{r}^{\mathrm{T}}C(t)\vec{r}}{|\vec{r}|^{2}}.
\label{eq:apx_costheta}
\end{equation}
The set $\mathcal{S}$ can then be written into
\begin{equation}
\mathcal{S} =\left\{\vec{r}~\Big|\cos\Theta=\frac{\vec{r}^{\mathrm{T}}C(t)\vec{r}}
{|\vec{r}|^{2}}, \exists t \right\}.
\end{equation}
Utilizing Eq.~(\ref{eq:apx_Ct_matrix}), Eq.~(\ref{eq:apx_costheta}) can be rewritten into
\begin{eqnarray}
\cos\theta &=& 1-\frac{1}{|\vec{r}|^2}\sum^{N-1}_{n=1}\sum^{n-1}_{i=0}
\left(1-\cos\left[(E_n-E_i)t\right]\right) \nonumber \\
& & \times \left(r^{2}_{n^{2}+2i-1}+r^{2}_{n^2+2i}\right),
\end{eqnarray}
where $r_{i}$ is the $i$th element of $\vec{r}$, which directly gives
\begin{eqnarray}
\mathcal{S} &=& \Bigg\{\vec{r}~\Big|1-\cos\Theta=\frac{1}{|\vec{r}|^2}\sum^{N-1}_{n=1}
\sum^{n-1}_{i=0}\left(1-\cos\left[(E_n-E_i)t\right]\right) \nonumber \\
& & \times \left(r^{2}_{n^{2}+2i-1}+r^{2}_{n^2+2i}\right), \exists t \Bigg\}.
\end{eqnarray}
This is a general expression of $\mathcal{S}$ for time-independent Hamiltonians
under unitary evolution.

\section{The QSL for time-independent Hamiltonians under unitary evolution}
\label{sec:apx_Nlevel}

\subsection{Proof with the assistance of $\mathcal{S}$}

The calculation is to utilize the set $\mathcal{S}$, in which all
states satisfy the equation
\begin{eqnarray}
1-\cos\Theta &=& \frac{1}{|\vec{r}|^2}\sum^{N-1}_{n=1}
\sum^{n-1}_{i=0}\left(1-\cos\left[(E_n-E_i)t\right]\right) \nonumber \\
& & \times \left(r^{2}_{n^{2}+2i-1}+r^{2}_{n^2+2i}\right). \label{eq:apx_SSS}
\end{eqnarray}
According to the definition, the operational definition of the QSL is the minimum
time satisfying this equation. Now define
\begin{eqnarray}
f(t) &:=& \frac{1}{|\vec{r}|^{2}} \sum_{n=1}^{N-1}\sum_{i=0}^{n-1}
\left(1-\cos \left[\left(E_{n}-E_{i}\right) t\right]\right) \nonumber \\
& & \times \left(r_{n^{2}+2 i-1}^{2}+r_{n^{2}+2 i}\right).
\end{eqnarray}
Its derivative on $t$ is
\begin{eqnarray}
\frac{\partial f}{\partial t} &=& \frac{1}{|\vec{r}|^{2}} \sum_{n=1}^{N-1}
\sum_{i=0}^{n-1}(E_{n}-E_{i})\sin [(E_{n}-E_{i}) t] \nonumber \\
& & \times (r_{n^{2}+2 i-1}^{2}+r_{n^{2}+2 i}). \label{eq:apx_dft}
\end{eqnarray}

The proof contains two steps: (1) We first prove that $f(\tau)$ is in the first
monotonic increasing regime of $f(t)$. To do that, we need to prove
$\frac{\partial f}{\partial t}\big|_{t=\tau}\geq 0$. The fact that
\begin{equation}
(E_{n}-E_{i})\tau = \frac{E_n-E_i}{E_{N-1}-E_0}\Theta \leq \Theta \leq \pi
\end{equation}
means the sine term in Eq.~(\ref{eq:apx_dft}) is non-negative; at the same
time, $E_{n}-E_{i}$ is also non-negative; thus, one can immediately obtain
$\frac{\partial f}{\partial t}\big|_{t=\tau}\geq 0$. The same result can be
obtained for any time $t\leq \tau$, indicating that $f(t)$ is a monotonic
increasing function in the regime $[0,\tau]$.

(2) Next we compare the values of $f(\tau)$ and $1-\cos\Theta$. Due to the equation
$1-\cos \left[\left(E_{n}-E_{i}\right) \tau\right] \leq 1-\cos \Theta$, one can have
\begin{eqnarray}
& & \sum_{n=1}^{N-1}\sum_{i=0}^{n-1}(1-\cos \left[\left(E_{n}-E_{i}\right)\tau\right])
\frac{r_{n^{2}+2 i-1}^{2}+r_{n^{2}+2 i}}{|\vec{r}|^{2}} \nonumber \\
& \leq & (1-\cos\Theta)\sum_{n=1}^{N-1}\sum_{i=0}^{n-1}
\frac{r_{n^{2}+2 i-1}^{2}+r_{n^{2}+2 i}}{|\vec{r}|^{2}} \nonumber \\
& \leq & (1-\cos\Theta),
\end{eqnarray}
which leads to
\begin{equation}
f(\tau) \leq 1-\cos \Theta.
\end{equation}

%========================Figure==========================
\begin{figure}[tp]
\centering
\includegraphics[width=8cm]{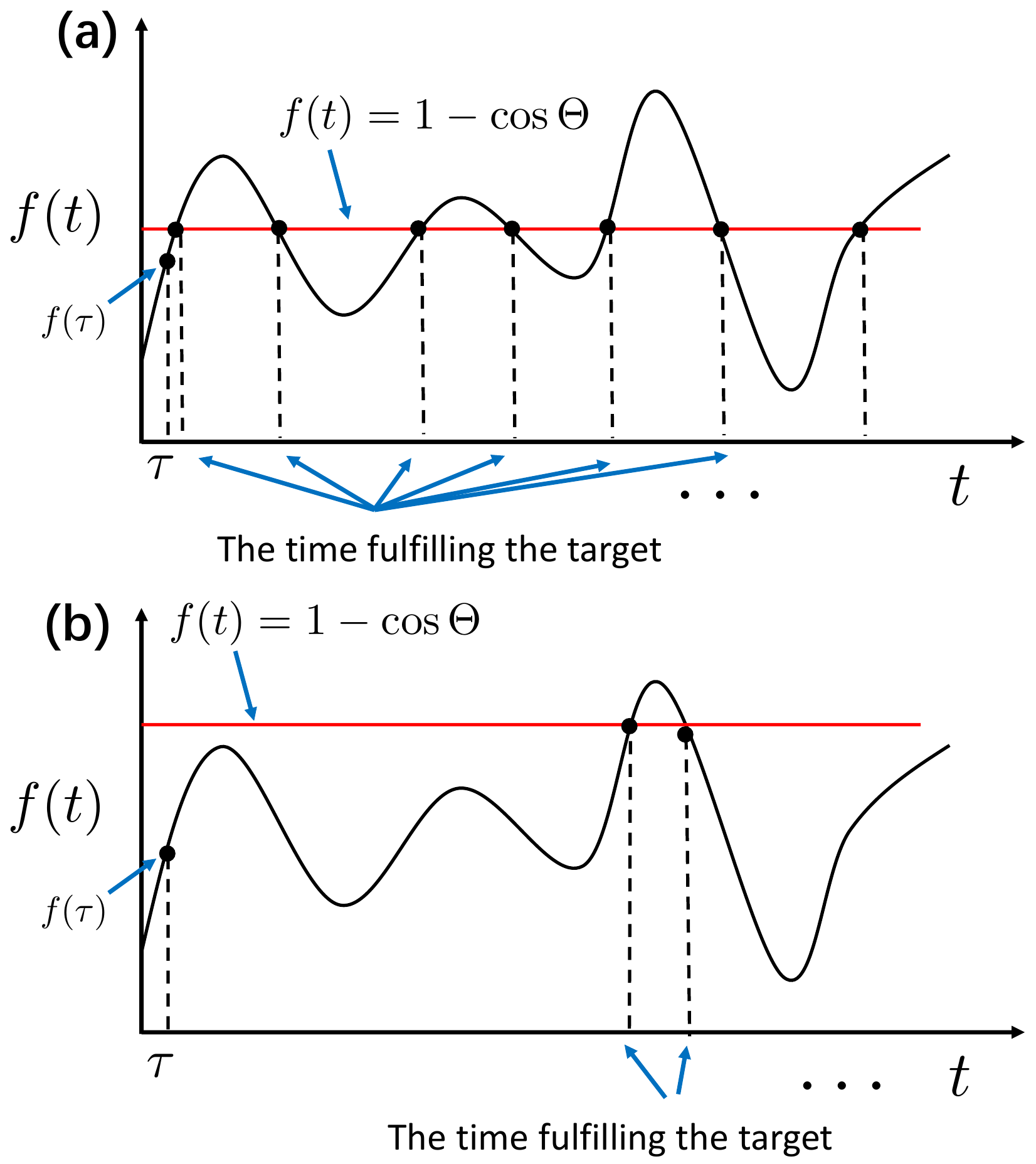}
\caption{(Color online) Schematic of $f(t)$ as a function of $t$. The red
line represents the value $1-\cos\Theta$. The crossover points between
two lines are the times at which the states reach the target angle.}
\label{fig:crossover_demo}
\end{figure}
%========================================================

In the case that the first crossover point between $f(t)$ and $1-\cos\Theta$ is in the
first monotonic increasing regime, as shown in Fig.~\ref{fig:crossover_demo}(a),
$t\geq \tau$ because $f(\tau) \leq 1-\cos \Theta$. In the case that the
first crossover point is not in the first monotonic increasing regime, as
shown in Fig.~\ref{fig:crossover_demo}(b), $t$ is also always larger than
$\tau$ since $\tau$ is always in the first monotonic regime. The result
$t\geq\tau$ is then proved.~~~~$\square$

The set $\mathcal{S}$ is worth studying. Denote $\mathcal{S}^{(km)}$ as
a subset of $\mathcal{S}$ in which all states satisfy
$r^{2}_{n^{2}+2i-1}+r^{2}_{n^{2}+2i}\neq 0$ for $n=k$ and $i=m$ and
$r^{2}_{n^{2}+2i-1}+r^{2}_{n^{2}+2i}=0$ for all the other subscripts.
For the set $\mathcal{S}^{(km)}$, the solution of $\tau$ satisfying
Eq.~(\ref{eq:apx_SSS}) is $t=\frac{\Theta}{E_k-E_m}$. Next, consider another set
$\mathcal{S}^{(km,lh)}\subset\mathcal{S}^{(km)}
\subset\mathcal{S}$, in which all states satisfy
$r^{2}_{n^{2}+2i-1}+r^{2}_{n^{2}+2i}\neq 0$
for both $n=k, i=m$ and $n=l, i=h$ and zero for all other subscripts.
It is obvious that $\mathcal{S}^{(km,lh)}\subset\mathcal{S}^{(lh)}$.
Utilizing the same strategy as we used above, it can be proved that 
the time given by $\mathcal{S}^{(km,lh)}$ is larger than
$\min\{\frac{\Theta}{E_k-E_m},\frac{\Theta}{E_l-E_h}\}$. In this way,
one can conclude that the time given by $\mathcal{S}^{(km,lh,\cdots)}$
is bounded by $\min\{\frac{\Theta}{E_k-E_m},\frac{\Theta}{E_l-E_h},\cdots\}$.

The states that can attain $\tau$ need to satisfy $r^2_{N^2-2N}+r^2_{N^2-2N+1}\neq 0$,
which is
\begin{equation}
\sum^{N-1}_{i=0}\frac{1}{N}|E_i\rangle\langle E_i|+\xi|E_0\rangle\langle E_{\rm max}|
+\xi^{*}|E_{\rm max}\rangle\langle E_0|,
\end{equation}
where $\xi=\sqrt{\frac{N-1}{2N}}(r_{N^2-2N}-i r_{N^2-2N+1})$. The specific
matrix formula in the energy basis $\{|E_m\rangle\}$ is
\begin{equation}
\left(\begin{array}{ccccc}
\frac{1}{N} & 0 & \cdots & 0 & \xi\\
0 & \frac{1}{N} & 0 & \cdots & 0\\
\vdots & 0 & \ddots & 0 & \vdots\\
0 & \cdots & 0 & \frac{1}{N} & 0\\
\xi^{*} & 0 & \cdots & 0 & \frac{1}{N}
\end{array}\right).
\end{equation}
To be a positive semi-definite matrix, $\xi$ should satisfy $|\xi|\in(0,1/N]$.

\subsection{Proof from the optimization of $\tau_{\mathrm{B}}$}

The target angle for the QSL is defined in various ways. With respect to the Bloch
vector, an elegant theoretical tool was provided by Campaioli \emph{et al.}~\cite{Campaioli2018},
which is a state-dependent bound with the expression
\begin{equation}
\tau_{\mathrm{B}} = \frac{\Theta}{Q},
\end{equation}
where
\begin{equation}
Q = \frac{1}{t}\int_0^t \mathrm{d}t' \sqrt{\frac{2\mathrm{Tr}(\rho_{t'}^2 H^2
-\rho_{t'} H\rho_{t'} H)}{\mathrm{Tr}(\rho_{t'}^2)-1/N}}.
\end{equation}
In the energy eigenspace $\{|E_m\rangle\}$, one can see that
\begin{eqnarray}
& &  \mathrm{Tr}(\rho_{t'}^2 H^2) \nonumber \\
&=& \frac{1}{N^2}\sum_n E_n^2+\frac{N-1}{2N}\sum_{ijn} r_i(t') r_j(t') E_n^2\langle
E_n |\lambda_i \lambda_j |E_n\rangle  \nonumber \\
& & +\frac{2}{N^2}\sqrt{\frac{N(N - 1)}{2}} \sum_{in}r_i(t') E_n^2\langle E_n
|\lambda _i | E_n\rangle.
\end{eqnarray}
Inserting $\openone=\sum_m|E_m\rangle\langle E_m| $ in the equation above,
we obtain
\begin{eqnarray*}
& &  \mathrm{Tr}(\rho_{t'}^2 H^2)  \\
&=& \frac{1}{N^2}\sum_n E_n^2 + \frac{2}{N^2}\sqrt{\frac{N(N - 1)}{2}}
\sum_{in} r_i(t') E_n^2\langle E_n|\lambda _i|E_n\rangle  \\
& & + \frac{N - 1}{2N}\sum_{ijmn} r_i(t') r_j(t') E_n^2\langle E_n |
\lambda_i| E_m\rangle\langle E_m |\lambda _j | E_n \rangle.
\end{eqnarray*}
Next, since
\begin{eqnarray*}
& & \mathrm{Tr}(\rho_{t'} H\rho_{t'} H)  \\
&=& \frac{1}{N^2}\sum_n E_n^2+\frac{2}{N}\sqrt{\frac{N(N - 1)}{2N}}
\sum_{i,n} r_i(t') E_n^2 \langle E_n |\lambda _i |E_n\rangle  \\
& &+ \frac{N-1}{2N}\sum_{ijmn}r_i(t') r_j(t') E_n E_m\langle E_n |\lambda _i| E_m\rangle
\langle E_m |\lambda_j |E_n \rangle,
\end{eqnarray*}
one can have
\begin{eqnarray*}
& & \frac{2N}{N-1}\mathrm{Tr}(\rho_{t'}^2 H^2 - \rho_{t'} H \rho_{t'} H)  \\
&=& \sum_{ijnm} r_i(t') r_j(t') E_n (E_n-E_m)\langle E_n|\lambda_i|E_m\rangle
\langle E_m | \lambda_j |E_n\rangle  \\
&=& \frac{1}{2}\sum_{ijnm} r_i(t') r_j(t')(E_n-E_m)^2\langle E_n|\lambda_i|E_m\rangle
\langle E_m | \lambda_j |E_n\rangle  \\
&=& \frac{1}{2}\!\sum_{ijnm}\!\! r_i(t') r_j(t')(E_n\!-\!E_m)^2 \mathrm{Re}(\langle E_n|
\lambda_i|E_m\rangle\! \langle E_m | \lambda_j |E_n\rangle).
\end{eqnarray*}
In the meantime, $\mathrm{Tr}(\rho^2)-\frac{1}{N}=\frac{N - 1}{N}|\vec{r}|^2$.
$Q$ can then be finally obtained as
\begin{eqnarray}
Q &=& \frac{1}{t}\int_0^t \mathrm{d}t' \sqrt{\sum_{ijnm}\frac{r_i(t') r_j(t')}
{2|\vec{r}|^2}(E_n-E_m)^2} \nonumber \\
& & \times \sqrt{\mathrm{Re}(\langle E_n|\lambda_i|E_m\rangle\langle E_m |\lambda_j |E_n\rangle)}.
\end{eqnarray}
For three-level systems, we chose Gell-Mann matrices as the $\mathfrak{su}(3)$ generators.
The non-zero terms in the summation in above equation are those with $i=j$.
Through some algebra, the term in the square root can be expressed by
\begin{eqnarray}
& & \frac{1}{|\vec{r}|^2}\Big\{[r_0^2(t')+r_1^2(t')](E_1-E_0)^2
+[r_3^2(t')+r_4^2(t')] \nonumber \\
& & \times (E_2-E_0)^2+[r_5^2(t')+r_6^2(t')](E_2-E_1)^2 \Big\}.
\end{eqnarray}
The maximum $Q$ can then be obtained when $r_3^2(t')+r_4^2(t')=|\vec{r}|^2$,
which gives $Q_{\max}=E_2-E_0$, and $\tau_{\mathrm{B}}$ reduces to $\tau$.

\section{One-dimensional transverse Ising model} \label{apx:Ising}

Explicit expressions can be derived in the continuum by replacing the discrete
sum over the set of quasimomenta by an integral, \emph{i.e.},
$\sum_k\rightarrow\frac{M}{2\pi}\int \mathrm{d}k$. The ground state energy then reads
\begin{eqnarray}
E_0&=&\frac{M}{2\pi}\int \omega_k \mathrm{d}k \nonumber\\
&=&-{\rm sgn}(h+1)\frac{2M(h+1)}{\pi}E\left(\frac{4h}{(h+1)^2}\right),
\end{eqnarray}
where $E(x)$ denotes the  complete elliptic integral of the second kind.
Similarly, it is found that
\begin{eqnarray*}
\delta\tau &=& \frac{\delta h \mathrm{sgn}(h+1)\pi \Theta/J}{8M h(h+1)^2 E^{2}
\left(\frac{4h}{(h+1)^2}\right)} \times  \\
& & \Bigg[\!(h+1)E\!\left(\frac{4h}{(h+1)^2}\right)\!+\!(h-1)
K\!\left(\frac{4h}{(h+1)^2}\right)\!\Bigg],
\end{eqnarray*}
where $K(x)$ denotes the complete elliptic integral of the first kind.
Note that $\delta\tau\propto M^{-1}$. In particular, in the neighborhood of
the critical point $h=1$,
\begin{eqnarray}
\delta\tau\approx\frac{\pi\delta h\Theta/J}{32 M}\left[5-3h-(h-1)
\log\left(\frac{h-1}{8}\right)\right].
\end{eqnarray}

\section{The QSL in two-level systems}
\label{sec:axp_twolevel}

%========================Figure==========================
\begin{figure}[tp]
\centering
\includegraphics[width=8cm]{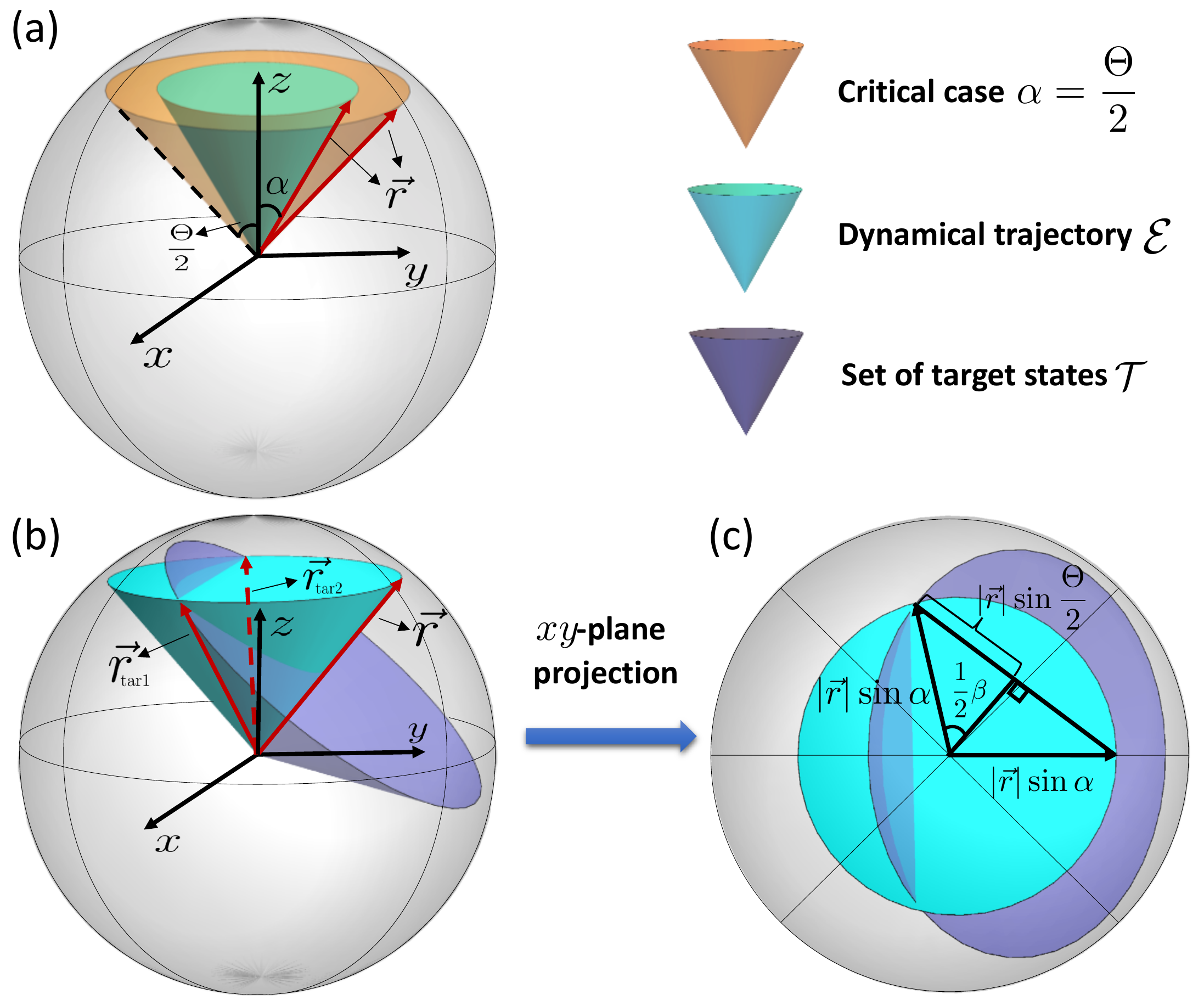}
\caption{(Color online) Schematic of three scenarios for the
calculation of the operational definition of the QSL in a qubit
system.  (a) The case $\alpha\leq\Theta/2$ ($\alpha$ is the
angle between the initial state and the $z$ axis). When
$\alpha<\Theta/2$, there exists no target state fulfilling
the angle $\Theta$. When $\alpha=\Theta/2$, only one target
state exists. (b) The case $\alpha >\Theta/2$. For this case,
two target states exist. (c) is the projection of initial and
target states on $xy$ plane.}
\label{Fig:qubit_appx}
\end{figure}
%========================================================

In this appendix we analyze two-level systems. The Hamiltonian of a two-level
system in the energy basis is $H=E_{0}|E_{0}\rangle\langle E_{0}|
+E_{1}|E_{1}\rangle\langle E_{1}|$, where $E_{0}$, $E_{1}$ are the energies and
$|E_{0}\rangle$, $|E_{1}\rangle$ are corresponding eigenstates. Define $\sigma_{z}$
as $\sigma_{z}:=|E_{1}\rangle\langle E_{1}|-|E_{0}\rangle\langle E_{0}|$, namely,
the Pauli matrix in basis $\{|E_{0}\rangle,|E_{1}\rangle\}$. With the Pauli
matrix, the Hamiltonian can be rewritten as $H=\frac{1}{2}(E_{0}+E_{1})\openone
+\frac{1}{2}(E_{1}-E_{0})\sigma_{z}$. The identity matrix $\openone$ commutes
with any operator, hence it has nothing to do with the evolution. Then the
Hamiltonian can be simplified into $\frac{1}{2}(E_{1}-E_{0})\sigma_{z}$.
In the Bloch representation, this means the evolution of any state is the rotation
of the corresponding Bloch vector about $z$ axis. A general vector in the Bloch sphere
can be expressed by
\begin{equation}
\vec{r}(\eta, \alpha,\varphi)=\eta(\sin\alpha\cos\varphi,\sin\alpha\sin\varphi,\cos\alpha),
\label{eq:qubit_bloch}
\end{equation}
where $\eta\in[0,1]$, $\alpha\in[0,\pi]$ and $\varphi\in[0,2\pi]$. For
an initial state $\vec{r}(\eta_{0}, \alpha_{0},\varphi_{0})$, the evolved state is
\begin{eqnarray*}
\vec{r}(t) &=& \eta\big(\sin\alpha\cos\varphi\cos(\omega t)-\sin\alpha\sin\varphi\sin(\omega t),  \\
& & \sin\alpha\cos\varphi\sin(\omega t)+\sin\alpha\sin\varphi\cos(\omega t),\cos\alpha\big).
\end{eqnarray*}
It can be seen in this equation that the period of the dynamics is
\begin{equation}
T=\frac{2\pi}{\omega}=\frac{2\pi}{E_{1}-E_{0}}.
\end{equation}
For two-level systems, utilizing Eq.~(\ref{eq:qubit_bloch}), the constrain in
$\mathcal{S}$ given in Proposition~\ref{prop:S_timeindependent} reduces to
\begin{equation}
\sin^2\left(\frac{\Theta}{2}\right)=\sin^2\left(\frac{\omega t}{2}\right)\sin^2\alpha.
\label{eq:constrain_qubit}
\end{equation}
The states with $\alpha=0$ does not evolve in this case, hence not in the set
$\mathcal{S}$. For $\alpha\neq 0$, the condition for $\alpha$ to make sure the
equation above has solutions for $t$ is
\begin{equation}
\sin^2\left(\frac{\Theta}{2}\right)\leq \sin^2\alpha,
\end{equation}
which is equivalent to
\begin{equation}
\alpha\in\left[\frac{\Theta}{2},\pi-\frac{\Theta}{2}\right].
\end{equation}
Furthermore, the minimum time under constrain~(\ref{eq:constrain_qubit}) is
reached when $\alpha$ is maximum, \emph{i.e.}, $\alpha=\pi/2$, which leads to
\begin{equation}
\tau=\frac{\Theta}{\omega}=\frac{\Theta}{E_1-E_0}.
\end{equation}
Corollary~\ref{corollary:qubit} is proved.
$\quad\quad\quad\quad\quad\quad\quad\quad\quad\quad\quad\quad\quad\quad\square$

To better understand the physics behind the QSL, we analyze the two-level systems
from a fully geometric perspective. For any specific initial state
$\vec{r}(\eta,\alpha,\varphi)$, the set of all states on the evolution trajectory
(denoted by $\mathcal{E}$) is
\begin{equation}
\mathcal{E} = \left\{\vec{r}(\eta,\alpha,\varphi)|\varphi\in[0,2\pi]\right\}.
\end{equation}
One may notice that the set of all target states for a specific initial state
(denoted by $\mathcal{T}$) here is a cone with the initial state as the axis
and $\Theta$ the central angle. For any state $\vec{r}(\eta, \alpha,\varphi)$,
the condition of $\vec{r}\in \mathcal{S}$ is that $\mathcal{E}$ and $\mathcal{T}$
have intersections.

In the case that $\alpha = \Theta/2$, $\mathcal{E}=\mathcal{T}=\{\vec{r}
(\eta,\frac{\Theta}{2},\varphi)|\varphi\in[0,2\pi] \}$ for any specific $\eta$,
as shown in the yellow cone in Fig.~\ref{Fig:qubit_appx}(a). The coincidence between
$\mathcal{E}$ and $\mathcal{T}$ means that all the states with $\eta\neq 0$ in $\mathcal{E}$
are in the set $\mathcal{S}$, \emph{i.e.,} $\mathcal{S}_{1}=\{\vec{r}(\eta,\frac{\Theta}{2},\varphi)|
\eta\in(0,1],\varphi\in[0,2\pi]\}\in\mathcal{S}$. Furthermore, it is easy to see that
for any specific state in this case, only one target state exists, i.e., the symmetrical
state with respect to the initial state about $z$-axis. It requires half of the period
to rotate the initial state to its symmetrical state, thus, the evolution time in this
scenario is
\begin{equation}
t = \frac{\pi}{\omega}=\frac{\pi}{E_{1}-E_{0}}.
\end{equation}

Next, for the case that $\alpha < \Theta/2$, all states within $\mathcal{E}$ (the blue cone in
Fig.~\ref{Fig:qubit_appx}(a)) fail to reach the target $\Theta$ since the largest angle between the initial
state and the evolved state is $2\alpha$, which is smaller than $\Theta$.
This means any state satisfying $\alpha<\Theta/2$ is not in the set $\mathcal{S}$.

For the case that $2\alpha > \Theta$, $\mathcal{E}$ (the blue cone in Fig.~\ref{Fig:qubit_appx}(b))
for any value of $\eta$ shares two vectors with $\mathcal{T}$ (the purple cone
in Fig.~\ref{Fig:qubit_appx}(b)), which means any state in this scenario has two
target states $\vec{r}_{\mathrm{tar1}}$ and $\vec{r}_{\mathrm{tar2}}$ on the
evolution trajectory. Thus, $\mathcal{S}_{2}=\{\vec{r}(\eta,\alpha,\varphi)|\eta\in (0,1],
\alpha>\frac{\Theta}{2},\varphi\in [0,2\pi]\}\in\mathcal{S}$.
Since the rotation is counterclockwise (looking against the $z$ axis), the evolution
time to $\vec{r}_{\mathrm{tar1}}$ is smaller than the one to $\vec{r}_{\mathrm{tar2}}$. To
calculate this evolution time, the angle between the projections of
$\vec{r}=\vec{r}(\eta,\alpha,\varphi)$ and $\vec{r}_{\mathrm{tar1}}$
on $xy$ plane (denoted as $\beta$) needs to be known. From Fig.~\ref{Fig:qubit_appx}(c),
it can be found that the length of the projection of $\vec{r}$ is $|\vec{r}|\sin\alpha$,
and the length between these two projections is $2|\vec{r}|\sin\left(\frac{\Theta}{2}\right)$.
Thus, the angle $\beta=2\arcsin\left(\frac{\sin\left(\frac{\Theta}{2}\right)}{\sin\alpha}\right)$,
which indicates that the evolution time is
\begin{equation}
t=\frac{2\pi}{E_{1}-E_{0}}\frac{\beta}{2\pi}=\frac{2}{E_{1}-E_{0}}
\arcsin\left(\frac{\sin\left(\frac{\Theta}{2}\right)}{\sin\alpha}\right).
\end{equation}
The minimum value of this evolution time is $\frac{\Theta}{E_{1}-E_{0}}$,
which is attained at $\alpha=\pi/2$. Combing the result obtained in the case of
$2\alpha=\Theta$, one can finally obtain $t \geq \frac{\Theta}{E_{1}-E_{0}}$,
and the set $\mathcal{S}=\mathcal{S}_{1}\cup\mathcal{S}_{2}$. The case
with $\pi-\alpha$ can be analyzed in the same way.

\section{The operational definition of the QSL in the Landau-Zener model} \label{apx:LZ}

The Hamiltonian of Landau-Zener model is
\begin{equation}
H=\Delta\sigma_{x}+vt\sigma_{z},
\end{equation}
where $\sigma_z=|1\rangle\langle 1|-|0\rangle\langle 0|$ with
$|\{|0\rangle,|1\rangle\}$ the computational basis. $\Delta$ and $v$ are two
time-independent parameters. For the case that $\Delta=0$,
$|0\rangle$ and $|1\rangle$ are the eigenstates of Hamiltonian.
The evolution operator for this Hamiltonian can then be calculated as
\begin{equation}
U=\exp\left(-\frac{i}{2}vt^{2}\sigma_{z}\right).
\end{equation}
In the following we will use the traditional notations $r_{x},r_{y},r_{z}$ as
the entries of Bloch vector instead of $r_0, r_1, r_2$. With above unitary
operator, the evolved Bloch vector can be calculated as below
\begin{eqnarray}
r_{x}\left(t\right) & = & \cos\left(vt^{2}\right)r_{x}-\sin\left(vt^{2}\right)r_{y}, \\
r_{y}\left(t\right) & = & \sin\left(vt^{2}\right)r_{x}+\cos\left(vt^{2}\right)r_{y}, \\
r_{z}\left(t\right) & = & r_{z}.
\end{eqnarray}
The angle between the initial and evolved states is then of the form
\begin{equation}
\cos\theta=\frac{\vec{r}\left(t\right)\cdot\vec{r}}{\left|\vec{r}\right|^{2}}
=\frac{\cos\left(vt^{2}\right)(r^{2}_{x}+r^{2}_y)+r_{z}^{2}}{|\vec{r}|^2},
\end{equation}
where $|\vec{r}|$ is the norm of $\vec{r}$. For the target angle $\Theta$, the
evolution time needs to satisfy the equation
\begin{equation}
\sin^{2}\left(\frac{vt^{2}}{2}\right)=\frac{|\vec{r}|^2}{|\vec{r}|^{2}-r^{2}_z}
\sin^{2}\left(\frac{\Theta}{2}\right).
\end{equation}
In the figure of $\sin^{2}\left(vt^{2}/2\right)$ as a function of $t$,
due to the fact that the first extremal value of $\sin^{2}\left(vt^{2}/2\right)$ is 1,
which is also the global maximum value, the first crossover point between it and the
line $\frac{|\vec{r}|^2}{|\vec{r}|^{2}-r^{2}_z}\sin^{2}\left(\frac{\Theta}{2}\right)$
is always in the first monotonic increasing regime, in which a smaller value of
$\frac{|\vec{r}|^2}{|\vec{r}|^{2}-r^{2}_z}\sin^{2}\left(\frac{\Theta}{2}\right)$
gives a smaller value of $t$. Therefore, the minimum time $\tau$ satisfying the
equation above is attained when $\frac{|\vec{r}|^2}{|\vec{r}|^{2}-r^{2}_z}$ is
minimum. Due to the fact that $\frac{|\vec{r}|^2}{|\vec{r}|^{2}-r^{2}_z}\geq 1$,
$\tau$ is of the form
\begin{equation}
\tau = \sqrt{\frac{\Theta}{v}},
\end{equation}
which is attained at $r_z = 0$, \emph{i.e.}, any state in the $xy$ plane.

%=======================Figure======================
\begin{figure}[tp]
\centering
\includegraphics[width=8cm]{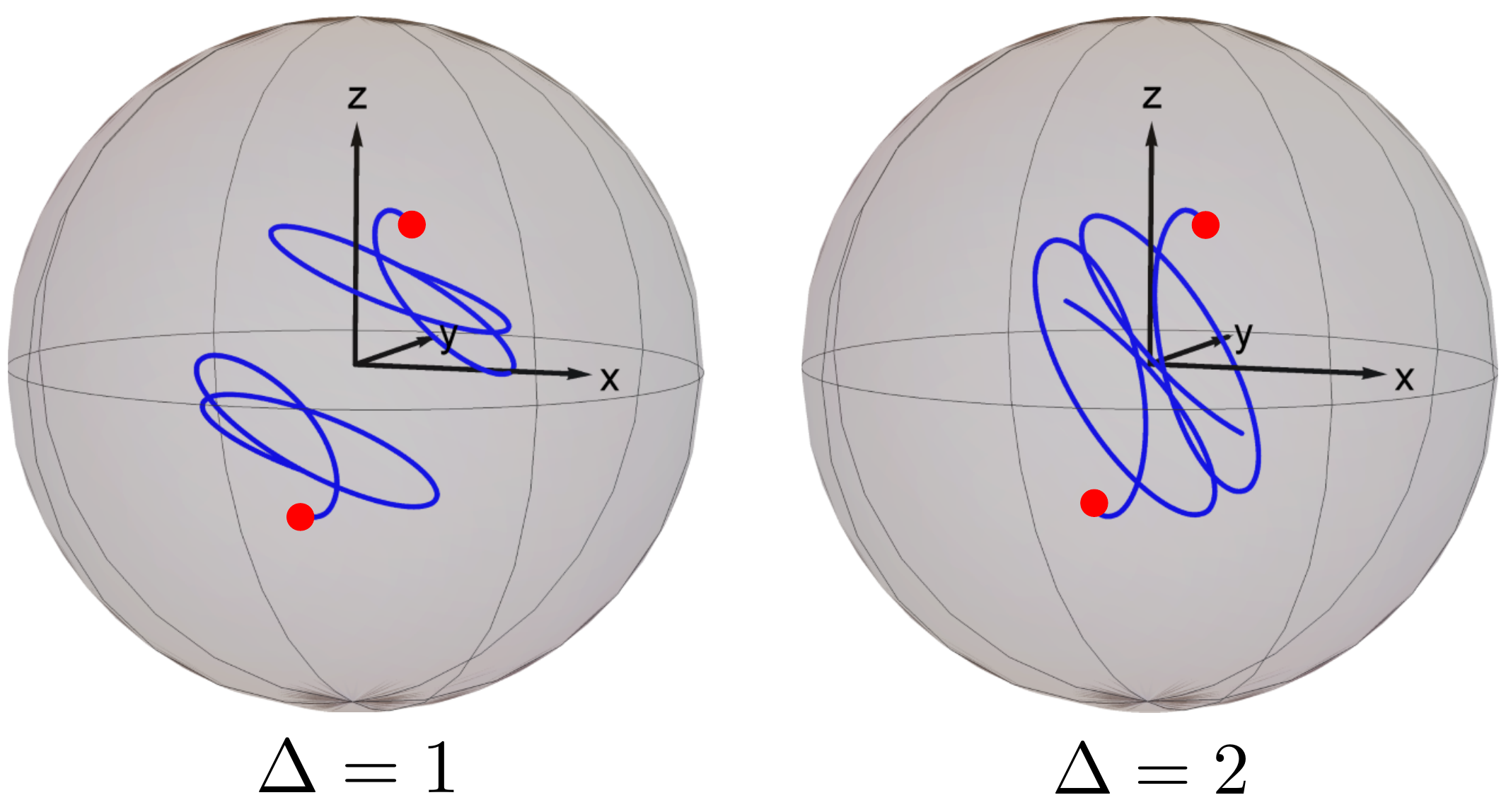}
\caption{(Color online) trajectories of center symmetric states (red dots) for
$\Delta=1.0$ (left) and $\Delta=2.0$ (right). The center symmetric states have
center symmetric trajectories, which is the reason for $\mathcal{S}$ to be
center symmetric.}
\label{fig:apx_LZ_trajectory}
\end{figure}
%===================================================

The set $\mathcal{S}$ (in Fig.~\ref{Fig:S_LZ}) for all values of $\Delta$ in
Landau-Zener model is center symmetric  about the original point, this is due
to the fact the trajectories of two center symmetric states are also center
symmetric, as shown in Fig.~\ref{fig:apx_LZ_trajectory}, which means the evolution
of the angles between the initial and evolved states are the same for these two
states. Therefore they can both reach the target angle simultaneously, which
is the reason why $\mathcal{S}$ is also center symmetric.

Next we calculate the bound $\tau_{\mathrm{B}}=\Theta/Q$~\cite{Campaioli2018}, where
\begin{equation}
Q = \frac{1}{t}\int_0^t \sqrt {\frac{2\mathrm{Tr}(\rho(t')^2 H^2
-\rho(t')H\rho(t')H)}{\mathrm{Tr}(\rho(t')^2)-1/2}} \mathrm{d}t'.
\end{equation}
Since $\mathrm{Tr}(\rho(t')^2 H^2)=\frac{1}{2}\left(\Delta^2+v^2 t'^2\right)
\left(1+|\vec{r}|^{2}\right)$, and
\begin{eqnarray}
\mathrm{Tr}(\rho(t')H\rho(t')H) &=& \frac{1}{2}\Delta^2 \left(1 -|\vec{r}|^2+2 r_x^2 \right)
+2\Delta vt' r_x r_z \nonumber \\
& & + \frac{1}{2} v^2 t'^2\left(1-|\vec{r}|^2+2 r_z^2\right).
\end{eqnarray}
Then one can have
\begin{eqnarray}
& & 2\mathrm{Tr}(\rho(t')^2 H^2) - 2\mathrm{Tr}(\rho(t')H\rho(t')H) \nonumber \\
& &= 2(\Delta^2+v^2 t'^2)|\vec{r}|^2 - 2(\Delta r_x+vt' r_z)^2.
\end{eqnarray}
In the meantime, $\mathrm{Tr}(\rho(t')^2)-\frac{1}{2}=\frac{1}{2}{|\vec{r}|^2}$,
which gives us the final expression of $Q$ in this case as
\begin{equation}
Q = \frac{2}{t} \int_0^t \mathrm{d}t' \sqrt{\Delta^2+v^2t'^2
-\frac{1}{|\vec{r}|^2}(\Delta r_x+vt' r_z )^2}.
\end{equation}
When $\Delta=0$, $r_z(t')=r_z$ is a constant, and the equation above reduces to
\begin{equation}
Q = vt\sqrt{1-\frac{r^2_z}{|\vec{r}|^2}}.
\end{equation}

\section{The QSL in open systems} \label{apx:sec_opensys}

\subsection{$\mathcal{S}$ for the general master equation}

For many quantum open systems, the dynamics is governed by the following
master equation
\begin{equation}
\partial_{t}\rho=-i\left[H,\rho\right]+\sum_{i}\gamma_{i}\left[L_{i}\rho L_{i}^{\dagger}
-\frac{1}{2}\{L_{i}^{\dagger}L_{i},\rho\}\right],
\end{equation}
where $\rho$ is a $N$-dimensional density matrix, and $L_{i}$ is
$i$th Lindblad operator depicting certain decay mode. Now we calculate
the set $\mathcal{S}$ for this dynamics. Substituting the Bloch representation
of $\rho$ into the equation above, one can obtain
\begin{eqnarray}
& & \sum_{k}\left(\partial_{t}r_{k}\right)\lambda_{k} \nonumber \\
&=&-i\sum_{k}r_{k}\left[H,\lambda_{k}\right]+\sum_{i}\gamma_{i}\sqrt{\frac{2}{N(N-1)}}
\left[L_{i},L_{i}^{\dagger}\right] \nonumber \\
& & +\sum_{i}\gamma_{i}\sum_{k}r_k\left(L_{i}\lambda_{k}L_{i}^{\dagger}
-\frac{1}{2}\left\{ L_{i}^{\dagger}L_{i},\lambda_{k}\right\} \right).
\end{eqnarray}
Recall that the $\mathfrak{su}(N)$ generators satisfy
\begin{align}
\left[\lambda_{k},\lambda_{l}\right] & =2i\sum_{m}\epsilon_{klm}\lambda_{m},\\
\left\{ \lambda_{k},\lambda_{l}\right\}  & =\frac{4}{N}\delta_{kl}\openone+2\sum_{m}\mu_{klm}\lambda_{m},
\end{align}
where $\epsilon_{klm}$ and $\mu_{klm}$ are some constants. Substituting
$\lambda_{l}$ into both sides of the equation above and taking the trace, one
can finally obtain the following equation
\begin{equation}
\partial_{t}\vec{r}=\mathcal{M}^{\mathrm{T}}\vec{r}+\vec{q},
\label{eq:apx_general_Bloch}
\end{equation}
which is an affine map with the entries of the coefficients
\begin{eqnarray}
\mathcal{M}_{kl} &=&
\sum_{i}\frac{\gamma_{i}}{2}\!\!\left[\mathrm{Tr}\!\left(L_{i}\lambda_{k}L_{i}^{\dagger}\lambda_{l}\right)
\!-\!\sum_{m}\mu_{klm}\mathrm{Tr}\!\left(L_{i}^{\dagger}L_{i}\lambda_{m}\right)\right]
\nonumber \\
& & +\!\sum_{m}\epsilon_{klm}\mathrm{Tr}\!\left(H\lambda_{m}\right)
-\frac{1}{N}\sum_{i}\gamma_{i}\delta_{kl}\mathrm{Tr}\!\left(L_{i}^{\dagger}L_{i}\right)\!\!,
\end{eqnarray}
and
\begin{equation}
q_{l}=\sum_{i}\frac{\gamma_{i}}{\sqrt{2N(N-1)}}\mathrm{Tr}
\left(\left[L_{i},L_{i}^{\dagger}\right]\lambda_{l}\right).
\end{equation}
In the case that $L_{i}$ can be decomposed with the generators, \emph{i.e.},
$L_{i}=e_{i,\mathrm{id}}\openone+\sum_{k}e_{i,k}\lambda_{k}$, the coefficients can
be rewritten as 
\begin{eqnarray}
\mathcal{M}_{kl} &=& \sum_{m}\epsilon_{klm}\left[\mathrm{Tr}\left(H\lambda_{m}\right)
+2\sum_{i}\gamma_{i}\mathrm{Im}(e_{i,\mathrm{id}}e^{*}_{i,m})\right] \nonumber \\
& & +\frac{2}{N}\sum_{i}\gamma_i\left(e_{i,k}e^{*}_{i,l}-\delta_{kl}\sum_{k'}|e_{i,k'}|^2\right) \nonumber \\
& & +\!\!\!\sum_{ik'mm'}\!\!\gamma_ie_{i,k'}e^{*}_{i,m'}\!\Big[\!(i\epsilon_{k'km}\!+\!\mu_{k'km})\!
(i\epsilon_{mm'l}\!+\!\mu_{mm'l}) \nonumber \\
& & -\mu_{klm}(i\epsilon_{mm'k'}+\mu_{mm'k'})\Big],
\end{eqnarray}
and
\begin{equation}
q_{l}=\sum_{ikk'}\frac{4\gamma_{i}\mathrm{Im}(e^{*}_{i,k} e_{i,k'})\epsilon_{kk'l}}
{\sqrt{2N(N-1)}}.
\end{equation}
For example, the coefficients for $N=2$ reduce to
\begin{eqnarray}
\mathcal{M}_{kl} &=& \sum_{m}\epsilon_{klm}\!\left[\mathrm{Tr}\left(H\lambda_{m}\right)
+2\sum_{i}\gamma_{i}\mathrm{Im}(e_{i,\mathrm{id}}e^{*}_{i,m})\right] \nonumber \\
& & +2\sum_{i}\gamma_i\!\!\left[\mathrm{Re}(e_{i,k}e^{*}_{i,l})-\delta_{kl}\sum_{k'}
|e_{i,k'}|^2\right]\!,
\end{eqnarray}
and $q_{l}=\sum_{ikk'}2\gamma_{i}\mathrm{Im}(e^{*}_{i,k} e_{i,k'})\epsilon_{kk'l}$.

In the case that $\mathcal{M}$ and $\vec{q}$ are time-independent, the solution
of Eq.~(\ref{eq:apx_general_Bloch}) is
\begin{equation}
\vec{r}(t)=e^{\mathcal{M}^{\mathrm{T}}t}\left(\vec{r}-\vec{l}\right)+\vec{l},
\end{equation}
where $\vec{l}$ satisfies $\mathcal{M}^{\mathrm{T}}\vec{l}=-\vec{q}$. The
inner product between $\vec{r}(t)$ and $\vec{r}$ then reads
\begin{equation}
\vec{r}(t)\cdot\vec{r}=\vec{r}^{\,\mathrm{T}}e^{\mathcal{M}^{\mathrm{T}}t}
(\vec{r}-\vec{l}~)+\vec{r}^{\,\mathrm{T}}\vec{l}.
\end{equation}
Therefore, the general expression of $\mathcal{S}$ for the above-mentioned
master equation is
\begin{equation}
\mathcal{S}=\left\{\vec{r}\,\Big|\cos\Theta=\frac{\vec{r}^{\,\mathrm{T}}
e^{\mathcal{M}^{\mathrm{T}}t}(\vec{r}-\vec{l}~)+\vec{r}^{\,\mathrm{T}}\vec{l}}
{|e^{\mathcal{M}^{\mathrm{T}}t}(\vec{r}-\vec{l}~)+\vec{l}\,||\vec{r}|},\exists t\right\}.
\end{equation}

\subsection{Spontaneous emission}

\emph{Calculation of $\mathcal{S}$}.
Here we show the analysis of the QSL for the dynamics
\begin{eqnarray}
\partial_{t}\rho &=&-i\left[H,\rho\right]+\gamma_{+} \left[\sigma_{+}\rho\sigma_{-}
-\frac{1}{2}\{\sigma_{-}\sigma_{+},\rho\}\right] \nonumber \\
& & +\gamma_{-} \left[\sigma_{-}\rho\sigma_{+}-\frac{1}{2}\{\sigma_{+}\sigma_{-},\rho\}\right],
\label{eq:apx_sigmaplus_deco}
\end{eqnarray}
where $\sigma_{\pm}=(\sigma_x\pm i\sigma_y)/2$ and $H=\omega_0\sigma_z/2$.
In the Bloch representation, $\mathcal{M}$ reads
\begin{equation*}
\mathcal{M}=\left(\begin{array}{ccc}
-\frac{1}{2}(\gamma_{+}+\gamma_{-}) & \omega_{0} & 0\\
-\omega_{0} & -\frac{1}{2}(\gamma_{+}+\gamma_{-}) & 0\\
0 & 0 & -(\gamma_{+}+\gamma_{-})
\end{array}\right),
\end{equation*}
and $\vec{q}=(0,0,\gamma_{+}-\gamma_{-})^{\mathrm{T}}$. Then the solution is
\begin{eqnarray}
r_{x}(t) & = & e^{-\frac{1}{2}(\gamma_{+}+\gamma_{-})t}
\left[\cos\left(\omega_{0}t\right) r_{x}(0)
-\sin\left(\omega_{0}t\right) r_{y}(0)\right], \nonumber \\
r_{y}(t) & = & e^{-\frac{1}{2}(\gamma_{+}+\gamma_{-})t}
\left[\cos\left(\omega_{0}t\right) r_{y}(0)
+\sin\left(\omega_{0}t\right) r_{x}(0)\right], \nonumber \\
r_{z}(t) & = & \frac{\gamma_{+}\!-\!\gamma_{-}}{\gamma_{+}
\!+\!\gamma_{-}}\!\!\left[1\!-\!e^{-(\gamma_{+}
+\gamma_{-})t}\right]\!\!+e^{-(\gamma_{+}+\gamma_{-})t} r_{z}(0).
\end{eqnarray}
Rewriting the initial state as
\begin{equation}
\vec{r}(0)=\eta\left(\sin\alpha\cos\varphi,\sin\alpha\sin\varphi,\cos\alpha\right),
\label{eq:apx_initial_r}
\end{equation}
the solutions reduce to
\begin{eqnarray}
r_{x}(t) &=& \eta e^{-(\gamma_{+}+\gamma_{-})t/2}\sin\alpha\cos\left(\omega_{0}t+\varphi\right) , \nonumber \\
r_{y}(t) &=& \eta e^{-(\gamma_{+}+\gamma_{-})t/2}\sin\alpha\sin\left(\omega_{0}t+\varphi\right) , \\
r_{z}(t) &=& \eta e^{-(\gamma_{+}+\gamma_{-})t} \cos\alpha+\frac{\gamma_{+}-\gamma_{-}}
{\gamma_{+}+\gamma_{-}}\left[1-e^{-(\gamma_{+}+\gamma_{-})t}\right]. \nonumber
\end{eqnarray}
The purity is of the form
\begin{eqnarray}
\left|\vec{r}(t)\right|^{2}
& \!=\! & \left\{\frac{\gamma_{+}\!-\!\gamma_{-}}{\gamma_{+}\!+\!\gamma_{-}}
\!\left[1-e^{-(\gamma_{+}+\gamma_{-})t}\right]
\!\!+\!\eta e^{-(\gamma_{+}+\gamma_{-})t}\!\cos\alpha\!\right\}^{2} \nonumber \\
& & +\eta^2 e^{-(\gamma_{+}+\gamma_{-})t}\sin^{2}\alpha.
\end{eqnarray}
In the mean time, the inner product between the initial and evolved states is 
\begin{eqnarray}
& & \vec{r}(0)\cdot\vec{r}(t) \nonumber \\
&=& \eta^2 e^{-(\gamma_{+}+\gamma_{-})t} \cos^{2}\alpha
+\eta^2 e^{-\frac{1}{2}(\gamma_{+}+\gamma_{-})t}\sin^{2}\alpha\cos\left(\omega_{0}t\right) \nonumber \\
& & +\eta \frac{\gamma_{+}-\gamma_{-}}{\gamma_{+}+\gamma_{-}}
\left[1-e^{-(\gamma_{+}+\gamma_{-})t}\right] \cos\alpha.
\end{eqnarray}
Hence, $\mathcal{S}$ in this case can be expressed by
\begin{equation}
\mathcal{S}=\left\{\!\vec{r}(\eta,\alpha)\big|\!\cos\Theta=\frac{\sin^{2}\alpha
\cos\left(\omega_{0}t\right)+\cos\alpha \chi}{\sqrt{\sin^{2}\alpha+\chi^2}},
\exists t\!\right\}, \label{eq:apx_sigmaplus_S}
\end{equation}
in which
\begin{equation}
\chi=e^{-\frac{1}{2}\gamma_{\mathrm{f}}t} \cos\alpha+\frac{2\gamma_{\mathrm{d}}}
{\eta \gamma_{\mathrm{f}}}\sinh\left(\frac{1}{2}\gamma_{\mathrm{f}}t\right).
\end{equation}
with $\gamma_{\mathrm{f}}=\gamma_{+}+\gamma_{-}$ and $\gamma_{\mathrm{d}}=\gamma_{+}-\gamma_{-}$.

\emph{Markovian dynamics.} Now we consider the case $\gamma_{+}=0$ and
$\gamma_{-}=\gamma$, which represents the dynamics of the spontaneous emission.
In this case, $\chi$ reduces to
\begin{equation}
\chi=e^{-\frac{1}{2}\gamma t}\cos\alpha-\frac{2}{\eta}\sinh\left(\frac{1}{2}\gamma t\right).
\end{equation}
Now we assume $\eta$ is very small (in the following we will use $\delta\eta$ instead)
and the time to reach the target angle could also be very small. For a very small
$\gamma t$, $\chi$ approximates to
\begin{equation}
\chi\approx\cos\alpha-\frac{\gamma t}{\delta\eta},
\end{equation}
with which the constrain in Eq.~(\ref{eq:apx_sigmaplus_S}) reduces to
\begin{eqnarray}
\cos\Theta &=& \frac{1-\frac{\gamma t}{\delta\eta}\cos\alpha}{\sqrt{1-\frac{2\gamma t}
{\delta\eta}\cos\alpha+\frac{\gamma^2 t^2}{(\delta\eta)^2}}} \nonumber \\
&=& \frac{1-\frac{\gamma t}{\delta\eta}\cos\alpha}{\sqrt{\left(1-\frac{\gamma t}
{\delta\eta}\cos\alpha\right)^2+\frac{\gamma^2 t^2}{(\delta\eta)^2}\sin^2\alpha}}.
\end{eqnarray}
Considering the case that $\Theta\in(0,\pi/2)$, the equation above is equivalent to
\begin{equation}
\cot\Theta=\frac{1-\frac{\gamma t}{\delta\eta}\cos\alpha}{\frac{\gamma t}
{\delta\eta}\sin\alpha},
\end{equation}
which can be rewritten as 
\begin{equation}
\sin\alpha\cot\Theta+\cos\alpha=\frac{\delta\eta}{\gamma t}.
\end{equation}
For a fixed $\delta\eta$, the minimum time can be obtained when the left-hand term is
maximum. Using the derivative of left-hand term with respect to $\alpha$
$\cos\alpha\cot\Theta-\sin\alpha$, one can immediately find out that the maximum
value is obtained when $\cot\Theta=\tan\alpha$, \emph{i.e.},
\begin{equation}
\alpha=\frac{\pi}{2}-\Theta.
\end{equation}
With this optimal initial state, $\tau$ reads
\begin{equation}
\tau = \frac{\delta\eta}{\gamma}\sin\Theta.
\end{equation}
A remarkable fact here is that $\tau$ is propositional to $\delta\eta$, which means
mixed initial states can provide a smaller $\tau$ than pure states.

\emph{non-Markovian dynamics.} This model ($\gamma_{+}=0$, $\gamma_{-}=\gamma$)
can also reveal the non-Markovian dynamics of damped Jaynes-Cummings models,
in which $\gamma=\gamma(t)$ is a time-dependent decay rate. Utilizing an effective
Lorentzian spectral density
\begin{equation}
J(\omega)=\frac{1}{2\pi}\frac{\gamma_0\lambda}{(\omega_0-\omega)^2+\lambda^2},
\end{equation}
$\gamma(t)$ can be analytically obtained as~\cite{Deffner2013}
\begin{equation}
\gamma(t)=\frac{8\gamma_0\lambda\sinh\left(\frac{1}{2}dt\right)}
{d\cosh\left(\frac{1}{2}dt\right)+\lambda\sinh\left(\frac{1}{2}dt\right)},
\end{equation}
where $d=\sqrt{\lambda^2-2\gamma_0\lambda}$. In this case, the entries of Bloch
vector read
\begin{eqnarray*}
r_x(t) &=& \eta  e^{-\frac{1}{2}\mathrm{Re}(\Gamma)} \sin\alpha
\cos\left(\frac{1}{2}\mathrm{Im}(\Gamma)+\omega_0 t+\varphi\right), \\
r_y(t) &=& \eta  e^{-\frac{1}{2}\mathrm{Re}(\Gamma)} \sin\alpha
\sin\left(\frac{1}{2}\mathrm{Im}(\Gamma)+\omega_0 t+\varphi\right), \\
r_z(t) &=& \eta \cos\alpha e^{-\mathrm{Re}(\Gamma)}-(1-e^{-\mathrm{Re}(\Gamma)}),
\end{eqnarray*}
where $\Gamma = \int^t_0\gamma(t')\mathrm{d}t'$ and $\mathrm{Re}(\cdot)$,
$\mathrm{Im}(\cdot)$ are the real and imaginary parts. With these expressions,
the norm square of $\vec{r}$ can be calculated as
\begin{equation}
|\vec r(t)|^2 = \eta^2 e^{-\mathrm{\rm Re}(\Gamma)}\sin^2\alpha
+\left[(1+\eta \cos \alpha)e^{-\mathop{\rm Re}(\Gamma)}-1\right]^2.
\end{equation}
In the mean time,
\begin{eqnarray*}
\vec{r}\cdot\vec{r}(t)&=& \eta^2 e^{-\frac{1}{2} \mathrm{Re}(\Gamma)}
\Bigg\{ \sin^2\alpha \cos\left(\frac{1}{2}\mathop{\rm Im}(\Gamma)+\omega_0 t\right) \\
& & +\cos\alpha \left[e^{-\frac{1}{2}\mathrm{Re}(\Gamma)}\cos\alpha-\frac{2}{\eta}
\sinh\left(\frac{1}{2}\mathrm{Re}(\Gamma)\right)\right] \Bigg\},
\end{eqnarray*}
which directly gives the set $\mathcal{S}$ as
\begin{equation}
\mathcal{S}\!=\!\left\{\vec{r}~\Big|\cos\Theta=
\frac{\sin^2\alpha\cos\left(\frac{1}{2}\mathrm{Im}(\Gamma)+\omega_0 t\right)
+\cos\alpha \chi_1}{\sqrt{\sin^2\alpha + \chi_1^2}}\right\}\!\!,
\end{equation}
where
\begin{equation}
\chi_1 = e^{-\frac{1}{2}\mathrm{Re}(\Gamma)}\cos\alpha
-\frac{2}{\eta}\sinh\left(\frac{1}{2}\mathrm{Re}(\Gamma)\right).
\end{equation}

The numerical calculation suggests that similar to the Markovian dynamics,
a lousy purity in this case can also benefit the reduction of $\tau$. Since
$\tau$ is very small here, $\Gamma$ would also be very small in the case that
$\gamma_0$ is not too large. In this case,  $\chi_1$ approximates to
$\chi_1\approx \cos\alpha-\frac{\mathrm{Re}(\Gamma)}{\delta\eta}$, which makes
\begin{equation}
\cos\Theta=\frac{1-\frac{\mathrm{Re}(\Gamma)}{\delta\eta}\cos\alpha}{\sqrt{\left(1-\frac{\mathrm{Re}(\Gamma)}
{\delta\eta}\cos\alpha\right)^2+\frac{\mathrm{Re}^2(\Gamma)}{(\delta\eta)^2}\sin^2\alpha}}.
\end{equation}
We also consider the case that $\Theta\in(0,\pi/2)$, the equation above equals to
\begin{equation}
\cot\Theta=\frac{1-\frac{\mathrm{Re}(\Gamma)}{\delta\eta}\cos\alpha}
{\frac{\mathrm{Re}(\Gamma)}{\delta\eta}\sin\alpha},
\end{equation}
which can be rewritten as 
\begin{equation}
\sin\alpha\cot\Theta+\cos\alpha=\frac{\delta\eta}{\mathrm{Re}(\Gamma)}.
\end{equation}
For a fixed $\delta\eta$, the minimum time can be obtained when the left-hand term is
maximum, which is the same as the Markovian case, \emph{i.e.}, $\alpha=\frac{\pi}{2}-\Theta$.
With this optimal initial state, we have $\mathrm{Re}(\Gamma)=\delta\eta \sin\Theta$.
Recalling the definition of $\Gamma$, one can obtain
\begin{equation}
\int^\tau_0 \frac{8\gamma_0\lambda\sinh\left(\frac{1}{2}dt'\right)}
{d\cosh\left(\frac{1}{2}dt'\right)+\lambda\sinh\left(\frac{1}{2}dt'\right)}\mathrm{d}t'
=\delta\eta \sin\Theta,
\end{equation}
which can be further solved as
\begin{equation}
\left(1-\frac{\lambda}{d}\right)e^{-\frac{1}{2}(d+\lambda)\tau}
+\left(1+\frac{\lambda}{d}\right)e^{\frac{1}{2}(d-\lambda)\tau}
=2e^{-\frac{1}{8}\delta\eta\sin\Theta}.
\end{equation}

\subsection{Parallel Dephasing}

Now we consider the dephasing model, in which the dynamics can be written as
\begin{equation}
\partial_{t}\rho=-i\left[\frac{1}{2}\omega_{0}\sigma_{z},\rho\right]
+\frac{\gamma}{2}\left(\sigma_{z}\rho\sigma_{z}-\rho \right).
\end{equation}
In this case $\mathcal{M}$ reads
\begin{equation}
\mathcal{M}=\left(\begin{array}{ccc}
-2\gamma & \omega_{0} & 0\\
-\omega_{0} & -2\gamma & 0\\
0 & 0 & 0
\end{array}\right),
\end{equation}
and $\vec{q}$ is a zero vector. Then the dynamics of the Bloch vector reads
\begin{eqnarray}
r_{x}(t) & = & e^{-\gamma t}\left[\cos\left(\omega_{0}t\right) r_{x}(0)
-\sin\left(\omega_{0}t\right) r_{y}(0)\right], \nonumber \\
r_{y}(t) & = & e^{-\gamma t}\left[\cos\left(\omega_{0}t\right) r_{y}(0)
+\sin\left(\omega_{0}t\right) r_{x}(0)\right], \nonumber \\
r_{z}(t) & = & r_{z}(0).
\end{eqnarray}
Rewriting the initial state as Eq.~(\ref{eq:apx_initial_r}), the solutions reduce to
\begin{eqnarray}
r_x (t) & = & \eta e^{-\gamma t} \sin\alpha \cos(\omega_{0}t+\varphi), \nonumber \\
r_y (t) & = & \eta e^{-\gamma t} \sin\alpha \sin(\omega_{0}t+\varphi), \nonumber \\
r_z (t) & = & \eta\cos\alpha.
\end{eqnarray}

Since the purity $\mathrm{Tr}(\rho^2)=\eta^2 \left(e^{-2 \gamma t}
\sin^{2}\alpha+\cos^{2}\alpha \right)$, the inner product between the initial
and evolved states is
\begin{equation}
\vec{r}(0)\cdot \vec{r}(t)=\eta^2 \left[e^{-\gamma t}
\cos\left(\omega_{0}t\right)\sin^{2}\alpha+\cos^{2}\alpha\right].
\end{equation}
Hence, $\mathcal{S}$ is of the form
\begin{equation}
\mathcal{S}=\left\{\vec{r}(\alpha)\Big|\!\cos\Theta\!=\!
\frac{1-\left[1-e^{-\gamma t}\cos(\omega_{0}t)\right]\sin^{2}\alpha}
{\sqrt{1-(1-e^{-2 \gamma t})\sin^{2}\alpha}}, \exists t \right\}.
\label{eq:apx_Sgamma2}
\end{equation}

To provide the regime of $\alpha$ in $\mathcal{S}$, we need to solve
$\sin^2\alpha$ in the constraint condition in Eq.~(\ref{eq:apx_Sgamma2}). Rewrite it as
\begin{equation}
x^2_1 y^2+(\cos^2\Theta x_2-2x_1)y+\sin^2\Theta=0, \label{apx_dephasing_square}
\end{equation}
where $y=\sin^2\alpha$, $x_1=1-e^{-\gamma t}\cos(\omega_0 t)$ and $x_2=1-e^{-2\gamma t}$.
The general solution for the equation above is
\begin{equation*}
y_{\pm}=\frac{1}{x_1}-\cos^2\Theta \frac{x_2}{2x^2_1}\pm\frac{\cos\Theta}{2x^2_1}
\sqrt{\cos^2\Theta x^2_2+4x^2_1-4x_1 x_2}.
\end{equation*}
To know if the expression of $y_{\pm}$ is exactly equivalent to Eq.~(\ref{eq:apx_Sgamma2})
(since Eq.~(\ref{apx_dephasing_square}) may bring extra solutions), the sign of
\begin{equation*}
1-x_1 y_{\pm}=\frac{\cos\Theta}{2x_1}\left(\cos\Theta x_2\mp\sqrt{\cos^2\Theta x^2_2
+4x^2_1-4x_1 x_2}\right)
\end{equation*}
needs to be checked to see if it coincides with $\cos\Theta$. In the case that
$\cos\Theta \geq 0$, $1-x_1 y_{-}$ is always positive and $1-x_1 y_{+}$ is only
positive when $\cos(\omega_0 t)>e^{-\gamma t}$, which means Eq.~(\ref{eq:apx_Sgamma2})
for $\cos\Theta\geq 0$ is actually equivalent to
\begin{equation}
\sin^{2}\alpha=\begin{cases}
y_{-}, & \text{for}~\cos(\omega_0 t) \leq e^{-\gamma t}, \\
y_{\pm}, & \text{for}~\cos(\omega_0 t) > e^{-\gamma t}.
\label{eq:apx_dephasing_sol1}
\end{cases}
\end{equation}
Using a similar analysis, one can see that for $\cos\Theta < 0$, Eq.~(\ref{eq:apx_Sgamma2})
is equivalent to
\begin{equation}
\sin^2\alpha=y_{-} \label{eq:apx_dephasing_sol2}
\end{equation}
when $\cos(\omega_0 t) \leq e^{-\gamma t}$ and no solution exists for other values of $t$.

Now we discuss the existence of solutions for $t$ using Eqs.~(\ref{eq:apx_dephasing_sol1})
and~(\ref{eq:apx_dephasing_sol2}) instead of Eq.~(\ref{eq:apx_Sgamma2}).
The solutions for $t$ exist only when $y_{\pm}$ is real and within
the regime $(0,1]$ for some values of $t$. The requirement for real solutions is
$x_{2}^{2}\cos^{2}\Theta-4x_{1}x_{2}+4x_{1}^{2}>0$, which cannot always be satisfied
for any value of $t$. When $t\rightarrow 0$, $x_{2}^{2}\cos^{2}\Theta-4x_{1}x_{2}+4x_{1}^{2}$
reduces to $4\gamma^2t^2(\cos^2\Theta-1)<0$, indicating that no state can
fulfill the target angle in an extremely small time. Furthermore, when
$t\rightarrow\infty$, $x_{2}^{2}\cos^{2}\Theta-4x_{1}x_{2}+4x_{1}^{2}$ reduces
to $\cos^2\Theta>0$. Therefore, the solution of time must be larger than
the time ($t_{\mathrm{c}}$) that first let $x_{2}^{2}\cos^{2}\Theta-4x_{1}x_{2}+4x_{1}^{2}$
be zero. Around the time $t_{\mathrm{c}}$, $y_{\pm}$ reduces to
\begin{eqnarray}
y_{\pm} &=& \frac{1}{x_1(t_{\mathrm{c}})}\left(1-\frac{x_2(t_{\mathrm{c}})}
{2x_1(t_{\mathrm{c}})}\cos^2\Theta\right) \nonumber \\
&\approx& \frac{2}{x_2(t_{\mathrm{c}})}-\frac{1}{x_1(t_{\mathrm{c}})}.
\end{eqnarray}
For a not very large $\gamma$, $t_{\mathrm{c}}$ always satisfies
$\cos(\omega_0 t_{\mathrm{c}})<e^{-\gamma t_{\mathrm{c}}}$, which immediately
gives $x_1(t_{\mathrm{c}})>x_2(t_{\mathrm{c}})$, then one can see that
\begin{equation}
y_{\pm}\geq \frac{1}{x_2(t_{\mathrm{c}})} \geq 1.
\end{equation}
Furthermore, in the same regime that $\cos(\omega_0 t)\leq e^{-\gamma t}$ can
be satisfied, $x_1(t)>x_2(t)$ always holds, which gives
\begin{eqnarray}
y_{-}\geq \frac{1}{x_1}-\cos^2\Theta \frac{x_2}{x^2_1}
\end{eqnarray}
for $\cos\Theta\geq 0$ and $y_{-}\geq 1/x_1$ for $\cos\Theta< 0$. These
two lower bounds can be both lower than 1 for a proper time. Hence, the value of
$y_{-}$ in this regime will continuously reduce to some value smaller than 1
from the time $t_{\mathrm{c}}$, which means the first cross point between $y_{-}$
and the regime $(0,1]$ has to be at 1, which corresponds to the shortest time solution
for Eqs.~(\ref{eq:apx_dephasing_sol1}) and~(\ref{eq:apx_dephasing_sol2}). At this
point, the constrain in Eq.~(\ref{eq:apx_Sgamma2}) reduces to $\cos\Theta=\cos(\omega_0 t)$,
which immediately gives the QSL as
\begin{equation}
\tau = \frac{\Theta}{\omega_0}.
\end{equation}

For example, in the case that $\Theta=\pi/2$, the constrain in
Eq.~(\ref{eq:apx_dephasing_sol1}) reduces to
\begin{equation}
\sin^2\alpha=\frac{1}{1-e^{-\gamma t}\cos(\omega_0 t)}.
\end{equation}
For a not very large $\gamma$, the smallest value of the right-hand side expression
is $[1+\exp(-\gamma \pi/\omega_0)]^{-1}$, which can be reached at $t=\pi/\omega_0$.
And it is obvious that its value can larger than 1; therefore, the regime
of $\sin^2\alpha$ in which the above equation has solutions for $t$ is
$\sin^2\alpha \in [(1+e^{-\frac{\gamma\pi}{\omega_0}})^{-1}, 1]$,
which directly leads to the regime of $\alpha$ in $\mathcal{S}$ as
\begin{equation*}
\alpha\!\in\!\left[\arcsin\!\left(\frac{1}{\sqrt{1+e^{-\frac{\gamma\pi}{\omega_0}}}}\right),
\pi\!-\!\arcsin\!\left(\frac{1}{\sqrt{1+e^{-\frac{\gamma\pi}{\omega_0}}}}\right)\right]\!,
\end{equation*}
and the QSL is $\tau =\pi/(2\omega_0)$.

\end{document}